\documentstyle[sprocl]{article}
\begin{document}
\title{Exact and Broken Symmetries in Particle Physics}
\author{R. D. Peccei}
\address{Department of Physics and Astronomy, UCLA, Los Angeles, CA 
90095-1547}
\maketitle
\begin{abstract}

In these lectures, I discuss the role of symmetries in particle physics. I begin by discussing global symmetries and show that they can be realized differently in nature, depending on whether or not the vacuum state is left invariant by the symmetry. I introduce next the notion of local symmetries and show how these symmetries can be implemented through the introduction of gauge fields. Using  the simple example of a spontaneously broken $U(1)$ symmetry, I discuss the Higgs mechanism showing that it provides a natural way for the gauge fields to acquire mass. Finally, I show how these concepts are used as the basis for the Standard Model of particle physics, ending with a brief description of some of the salient aspects of Quantum Chromodynamics and of the electroweak theory. 

\end{abstract}

\section{Introduction}

All experimental evidence points to the strong, weak and electromagnetic interactions
of hadrons (strongly interacting particles) and of leptons as being described
by a gauge theory, based on the group
\begin{equation}
G_{\rm SM} = SU(3)\times SU(2)\times U(1)~.
\end{equation}
The strong interaction theory--QCD--has as fundamental fermionic entities
a triplet of quarks, which feel the $SU(3)$ gauge interactions.  Both the
quarks and the leptons appear in nature in a repetitive fashion, in three
distinct families of doublets under the $SU(2)\times U(1)$ electroweak
group.  Although $G_{\rm SM}$ correctly
describes the symmetry of the fundamental interactions among quarks and
leptons, only $SU(3)$ is an exact symmetry of the theory.  The electroweak
group, in fact, suffers a spontaneous breakdown to $U(1)_{\rm em}$:
\begin{equation}
SU(2)\times U(1)\to U(1)_{\rm em}~.
\end{equation}      

In these lectures we will describe the fundamental
concepts upon which the theory for these interactions is built upon.
These are related to the way in which symmetries are realized in nature
and to the role of gauge fields in rendering theories invariant under local
transformations.  A crucial notion is that of a spontaneously broken
symmetry and the effect that this spontaneous breakdown has for the
spectrum of excitations in the theory.

\section{Global Symmetries in Field Theory}

The natural language for elementary particle physics is that of a quantum
field theory, where to each fundamental excitation one assigns a corresponding
quantum field.~\cite{Ramond}  Symmetries of nature are incorporated by constructing
Lagrangian densities, made up of these quantum fields, which have an action
\begin{equation}
W = \int d^4x {\cal{L}}
\end{equation}
explicitly invariant under the symmetry in question:
\begin{equation}
W\to W^\prime = W~.
\end{equation}

In what follows, I will consider only continuous symmetry transformations
based on some Lie group $G$.~\cite{Group}  Let me denote a generic quantum field by
$\chi_{\alpha}(x)$, which $x$ being the space-time location of the quantum field and
$\alpha$ being an (internal) index which runs over the possible components of
$\chi$.  [For instance, for a quark field which is a triplet of $SU(3)$ one
would have $q_\alpha(x)$, with $\alpha = 1,2,3$.]  If $a$ is one of the
operations of the symmetry group $G$ of transformations, and if the quantum
fields $\chi_\alpha$ are members of an (irreducible) multiplet, then under this
operation one has
\begin{equation}
\chi_\alpha(x) \stackrel{a}{\rightarrow} \chi^\prime_\alpha(x) =
{\cal{R}}_{\alpha\beta}(a) \chi_\beta(x)~.
\end{equation}
That is, under the transformation the field $\chi$ goes into a new field
$\chi^\prime$ whose components are linear combinations of the old
components.

Because, by assumption, the quantum fields $\chi_\alpha$ are members of a
multiplet under $G$, the matrices ${\cal{R}}(a)$ constitute a representation
matrix for the group $G$ and obey a characteristic composition property.
This follows from comparing the sequence of transformations
\begin{equation}
\chi_\alpha(x)\stackrel{a}{\rightarrow}\chi^\prime_\alpha(x)
\stackrel{a^\prime}{\rightarrow} \chi^{\prime\prime}_{\alpha}(x)
\end{equation}
to the direct transformation
\begin{equation}
\chi_\alpha(x)\stackrel{a^{\prime\prime}}{\rightarrow}
\chi^{\prime\prime}_{\alpha}(x)~.
\end{equation}
Hence, one finds
\begin{equation}
{\cal{R}}_{\alpha\beta}(a^\prime) {\cal{R}}_{\beta\gamma}(a) =
{\cal{R}}_{\alpha\gamma}(a^{\prime\prime})
\end{equation}

In the Hilbert space of the quantum field $\chi_\alpha(x)$ the transformation
(5) is induced by a {\bf unitary operator} $U(a)$, so that
\begin{equation}
U^{-1}(a)\chi_\alpha(x) U(a) = \chi^\prime_\alpha(x) =
{\cal{R}}_{\alpha\beta}(a)\chi_\beta(x)~.
\end{equation}
It is easy to see that the composition property (8) has its counterpart
in terms of the unitary operators $U$:
\begin{equation}
U(a) U(a^\prime) = U(a^{\prime\prime})~.
\end{equation}
Since we are considering continuous symmetry transformations, it suffices to
focus only on infinitesimal transformations $\delta a$, since finite
transformations can always be built up via (10) by (infinite) compounding.
A given Lie group is characterized by the number of parameters associated
with these infinitesimal transformations and, more specifically, by the
algebra obeyed by the operators connected to the distinct infinitesimal
parameters.

Let us write for an infinitesimal transformation
\begin{equation}
U(\delta a) = 1 + i\delta a_i G_i
\end{equation}
where the index $i$ runs over all the independent infinitesimal parameters
of the Lie group G [e.g. for the rotation group in 3 dimensions $O(3)$, $\delta a_i$
would describe the three independent rotations about the $x,y$ and $z$
axis].  The operators $G_i$ are called the group generators and the
composition property (10) implies a group algebra for the generators.
Without loss of generality the parameters $\delta a_i$ can be taken as real,
so that the $G_i$ are Hermitian.  They obey the Lie algebra:
\begin{equation}
[G_i,G_j] = i~c_{ijk}G_k~.
\end{equation}
The structure constants $c_{ijk}$ characterize the group $G$ and can be
chosen so as to be totally antisymmetric in $i,j$ and $k$.

Just as $U(\delta a)$ can be expanded in terms of the generators $G_i$, so
can the representation matrices ${\cal{R}}_{\alpha\beta}(\delta a)$.
One has, for an infinitesimal transformation
\begin{equation}
{\cal{R}}_{\alpha\beta}(\delta a) = \delta_{\alpha\beta} +
i\delta a_i(g_i)_{\alpha\beta}~.
\end{equation}
It is easy to show that the matrices $g_i$ furnish a representation for the
generators $G_i$ and so obey themselves Eq. (12).  To see this let us use
(13) and (11) in the defining equation (9). One has
\begin{equation}
(1-i\delta a_iG_i)\chi_\alpha(x)(1+i\delta a_iG_i) =
\chi_\alpha(x) + i\delta a_i(g_i)_{\alpha\beta}\chi_\beta
\end{equation}
which implies
\begin{equation}
[G_i,\chi_\alpha(x)] = -(g_i)_{\alpha\beta}\chi_\beta(x)~.
\end{equation}
This equation embodies succinctly how the quantum fields $\chi_\alpha$
transform under the group $G$, and will be repeatedly used in what follows.
By using (15) in the Jacobi identity
\begin{equation}
[G_i,[G_j,\chi_\alpha]] + [\chi_\alpha,[G_i,G_j]] +
[G_j,[\chi_\alpha,G_i]] = 0
\end{equation}
one readily sees that the matrices $g_i$ obey Eq. (12).

Let us explore the consequences of having a theory built out of the quantum
fields $\chi_\alpha$ which is invariant under the transformations of the
group $G$. As we shall see, the invariance of the action under $G$ implies the
existence of conserved currents and a set of constants of the motion, which
are nothing else but the generators $G_i$ of the group!  Since the Lagrangian
density ${\cal{L}}$ depends in general on $\chi_\alpha$ and its space-time
derivatives $\partial_\mu\chi_\alpha$, the invariance statement (4) implies
\begin{equation}
\int d^4x {\cal{L}}(\chi_\alpha,\partial_\mu\chi_\alpha) =
\int d^4x {\cal{L}}(\chi^\prime_\alpha,\partial_\mu\chi^\prime_\alpha)~.
\end{equation}
For $\chi^\prime_\alpha$ infinitesimally different from $\chi_\alpha$, the
stationarity of the action implies
\begin{eqnarray}
0 = \delta W &=& \int d^4x \left[\frac{\partial{\cal{L}}}{\partial\chi_\alpha}
\delta\chi_\alpha + \frac{\partial{\cal{L}}}{\partial\partial_\mu\chi_\alpha}
\delta\partial_\mu\chi_\alpha\right]  \\ \nonumber
&=& \int d^4x\left[\left\{\frac{\partial{\cal{L}}}{\partial\chi_\alpha} -
\partial_\mu\left(\frac{\partial{\cal{L}}}{\partial\partial_\mu\chi_\alpha}
\right)\right\} \delta\chi_\alpha \right.  \\ \nonumber
& &\mbox{}+\left.\partial_{\mu}\left[\frac{\partial{\cal{L}}}{\partial\partial_\mu\chi_\alpha}
\delta\chi_\alpha\right]\right]~.
\end{eqnarray}
The first term above in the curly brackets vanishes because of the Euler-Lagrange
equations of motion.  The second can be rewritten in terms of the generator
matrices $g_i$, since
\begin{equation}
\delta\chi_\alpha = \chi^\prime_\alpha - \chi_\alpha =
i~\delta a_i(g_i)_{\alpha\beta}\chi_\beta~.
\end{equation}
Hence

\begin{equation}
0 = \delta W = -\int d^4x\delta a_i \partial_\mu
\left[\frac{\partial{\cal{L}}}{\partial\partial_\mu\chi_\alpha}
\frac{1}{i}(g_i)_{\alpha\beta}\chi_\beta\right]~.
\end{equation}
Since the parameters $\delta a_i$ are independent, it follows that the
currents
\begin{equation}
J^\mu_i(x) = \frac{\partial{\cal{L}}}{\partial\partial_\mu\chi_\alpha(x)}
\frac{1}{i} (g_i)_{\alpha\beta}\chi_\beta(x)~,
\end{equation}
as a result of the symmetry, are conserved
\begin{equation}
\partial_\mu J^\mu_i(x) = 0~.
\end{equation}

Because of (22)--- if one assumes that the fields $\chi_\alpha$ drop off
sufficiently fast at spatial infinity--- there exists a set of constants of
the motion, given by the space integral of the $J_i^o$.  One has
\begin{equation}
Q_i = \int d^3x J^o_i(x)
\end{equation}
with
\begin{equation}
\frac{d}{dt} Q_i = 0~.
\end{equation}
It is easy to check--- and we shall do so below--- that the operators $Q_i$ are
precisely the generators $G_i$.  That is, they obey both Eqs. (12) and
(15).  If $H$ is the Hamiltonian of the theory, then Heisenberg's equation
of motion imply
\begin{equation}
[H,G_i] = 0
\end{equation}
which may be a more familiar way to express the invariance of the theory under
the transformations of the group $G$ (e.g. rotational invariance is expressed
via the vanishing of the commutator $[H,L_i] = 0$).

Let us verify that indeed
\begin{equation}
G_i\equiv Q_i = \int d^3x J_i^o = \int d^3x
\left[\frac{\partial{\cal{L}}}{\partial\partial_o\chi_\alpha}
\frac{1}{i} (g_i)_{\alpha\beta} \chi_\beta\right]
\end{equation}
acts as a generator is supposed to do.  For that, remark that the canonical
momentum conjugate to $\chi_\alpha$ is precisely~\cite{Ramond}
\begin{equation}
\pi_\alpha(x) = \frac{\partial{\cal{L}}}{\partial\partial_o\chi_\alpha(x)}
\end{equation}
and that (for bosonic fields) one has the equal time commutation relations
\begin{eqnarray}
\left.[\pi_\alpha(x),\chi_\beta(y)]\right|_{x^o=y^o}& =&\frac{1}{i}\delta^3(\vec{x}-\vec{y})\delta_{\alpha\beta} \\ \nonumber
\left.[\pi_\alpha(x),\pi_\beta(y)]\right|_{x^o=y^o}&=&\left.[\chi_\alpha(x), \chi_\beta(y)]\right|_{x^o=y^o} = 0~.
\end{eqnarray}
Then
\begin{equation}
G_i = \int d^3x \pi_\alpha(x)\frac{1}{i}(g_i)_{\alpha\beta}
\chi_\beta(x)~.
\end{equation}
Since $G_i$ is time-independent, in computing the commutator of $G_i$ with
$\chi_\gamma(y)$ one can set the time $x^o$ in (29) equal to $y^o$.
Using (28) it is then trivial to check that
\begin{eqnarray}
[G_i,\chi_\gamma(y)] &=& \int d^3x\left[\pi_\alpha(x)\frac{1}{i} (g_i)_{\alpha\beta}\chi_\beta(x),\chi_\gamma(y)\right]_{x^o=y^o} \nonumber \\
&=& -(g_i)_{\gamma\beta}\chi_\beta(y)
\end{eqnarray}
and
\begin{eqnarray}
[G_i,G_j] &=& \int d^3x d^3y\left[\pi_\alpha(x)\frac{1}{i}
(g_i)_{\alpha\beta}\chi_\beta(x),~\pi_\gamma(y)\frac{1}{i}(g_j)_{\gamma\delta}
\chi_\delta(y)\right]_{x^o=y^o} \nonumber \\
&=& \int d^3x \pi_\alpha(x)\left(\frac{1}{i}[g_i,g_j]\right)_{\alpha\beta}
\chi_\beta(x) \nonumber \\
&=& i~c_{ijk} \int d^3x \pi_\alpha(x)\frac{1}{i}(g_k)_{\alpha\beta}
\chi_\beta(x) = i~c_{ijk} G_k~.
\end{eqnarray}

Up to now in the discussion of symmetries I focussed on the transformation
properties of the quantum fields $\chi_\alpha(x)$.  What equation (9) says
is that under a group transformation the component fields $\chi_\alpha$
transform in a well-defined way.  The correspondence between quantum fields and
particles makes it natural to suppose that the quantum states associated with
the fields $\chi_\alpha(x)$ will transform in an analogous way.  Let me denote
the one-particle state associated with the field $\chi_\alpha$ by $|p;\alpha\rangle$,
where $p^\mu$ is the 4-momentum of the state and, since
these states are supposed to describe particles of a given mass, $p^2 = -m^2_\alpha$.  Then, 
corresponding to Eq. (9), one has
\begin{equation}
U^{-1}(a)|p;\alpha\rangle = {\cal{R}}_{\alpha\beta}(a)|p;\beta\rangle~.
\end{equation}
This equation can be used to deduce that all states of the multiplet
$|p;\alpha\rangle$ have the same mass.

Let $|p;\alpha\rangle_{\rm rest}$ denote the state corresponding to
4-momentum $p^\mu = (\vec 0,m_\alpha)$.  Then, by definition, the action of
the Hamiltonian on this state is just 
\begin{equation}
H|p;\left.\alpha\right\rangle_{\rm rest} = 
m_\alpha|p;\left.\alpha\right\rangle_{\rm rest}~.
\end{equation}
However, if the theory is invariant under the group $G$, so that $H$ commutes
with all the generators ( c. f. Eq. (25)) it follows also that
\begin{equation}
[H,U^{-1}(a)] = 0~.
\end{equation}
Applying this equation on the rest state proves our contention, since
\begin{eqnarray}
0 &=& [H,U^{-1}(a)]|p;\left.\alpha\right\rangle_{\rm rest} =
(HU^{-1}(a) - U^{-1}(a)H)|p;\left.\alpha\right\rangle_{\rm rest} \nonumber \\
&=& {\cal{R}}_{\alpha\beta}(a)(m_\beta-m_\alpha)|p;\left.
\beta\right\rangle_{\rm rest}~.
\end{eqnarray}
Because ${\cal{R}}_{\alpha\beta}(a)$ is arbitrary, it follows that
$m_\alpha = m_\beta$.  

One says that a symmetry is realized in a Wigner-Weyl
way if the invariance of the action under $G$ leads to the appearance in
nature of particle multiplets with the same mass.~\cite{WW}  A well known example of
an (approximate) Wigner-Weyl symmetry is strong isospin.  This approximate
global $SU(2)$ symmetry of the strong interaction leads to a nearly degenerate
nucleon doublet $(m_p\simeq m_n)$ and a pion triplet $(m_{\pi^+}=m_{\pi^-}
\simeq m_{\pi^o})$.  Remarkably, however, the Wigner-Weyl way is {\bf not}
the only way in which a symmetry can be realized in nature!

\section*{The Nambu-Goldstone Realization}

It is possible that the action is invariant under a symmetry group $G$ but
that the physical states of the theory show no trace of this symmetry.  This
happens in the case in which, although
\begin{equation}
[H,U^{-1}(a)] = 0~,
\end{equation}
the {\bf vacuum} state is not invariant under $G$.  Such symmetries are called
spontaneously broken, or realized in a Nambu-Goldstone way.~\cite{NG}

Eq. (32), which lead to the deduction that all states in a multiplet
$|p,\alpha\rangle$ have the same mass, can be derived from the transformation
properties of the quantum fields $\chi_\alpha$, provided one assumes that the
vacuum state is $G$ invariant:
\begin{equation}
U(a)|0\rangle = |0\rangle~.
\end{equation}
The one particle states $|p,\alpha\rangle$ are constructed by the action of
the (asymptotic) creation operators for the field $\chi_\alpha$.~\cite{LSZ} For a scalar field $\chi_\alpha(x)$ one writes in the usual way
\begin{equation}
\chi_\alpha(x) = \int \frac{d^3p}{(2\pi)^32p^o}
[e^{ipx}a_\alpha(p,t) + e^{-ipx}a^{\dagger}_\alpha(p,t)]~.
\end{equation}
Then, one has
\begin{eqnarray}
|p;\alpha\rangle &=& \lim_{t\to\pm\infty} a^{\dagger}_\alpha(p,t)|0\rangle \nonumber \\
&=& \lim_{x^o\to\pm\infty} \int d^3x e^{ipx}\frac{1}{i}
{\stackrel{\leftrightarrow}{\partial}}_o \chi_\alpha(x)|0\rangle~,
\end{eqnarray}
where
\begin{equation}
A{\stackrel{\leftrightarrow}{\partial}}_o B = A\partial_o B -
(\partial_oA)B~.
\end{equation}
Consider then, as in Eq. (32), the action of $U^{-1}(a)$ on the state
$|p;\alpha\rangle$
\begin{equation}
U^{-1}(a)|p,\alpha\rangle = \lim_{x^o\to\pm\infty}\int d^3x e^{ipx}
\frac{1}{i}{\stackrel{\leftrightarrow}{\partial}}_o U^{-1}(a)
\chi_\alpha(x)|0\rangle~.
\end{equation}
If (37) holds, one can write
\begin{eqnarray}
U^{-1}(a)\chi_\alpha(x)|0\rangle &=& U^{-1}(a) \chi_\alpha(x) U(a)|0\rangle
\nonumber \\
&=& {\cal{R}}_{\alpha\beta}(a)\chi_\beta(x)|0\rangle
\end{eqnarray}
which immediately establishes (32).  However, if the vacuum is {\bf not}
left invariant by a $G$-transformation--- i.e. if the vacuum state is degenerate
or not unique--- then even though the fields  $\chi_\alpha$ transform  according to some irreducible representation, there are
no longer degenerate multiplets in the spectrum.

When a symmetry is realized in a Nambu-Goldstone way, instead of having 
multiplets of particles with the same mass, there appear in the theory massless
excitations--- the so-called Goldstone bosons.  To see how these ensue consider
again the fields $\chi_\alpha$ and take the vacuum expectation value of
Eq. (15)
\begin{equation}
\langle 0|[G_i,\chi_\alpha(x)]|0\rangle = -(g_i)_{\alpha\beta}
\langle 0|\chi_\beta(x)|0\rangle~.
\end{equation}
If the vacuum is invariant under $G$ transformations it
follows  from Eq. (37) that
\begin{equation}
G_i|0\rangle = 0.
\end{equation}
It is immediate from (43) then that the vacuum expectation values of the fields $\chi_\alpha$ must vanish.
However, if (44) does {\bf not} hold, and $\chi_\alpha$ are scalar fields,
there is no argument why one cannot have
\begin{equation}
\langle 0|\chi_\alpha(x)|0\rangle \not= 0~.
\end{equation}
[If $\chi_\alpha$ correspond to fields with spin then the equivalent of
Eq. (43) for Lorentz transformations, along with the invariance of the vacuum
under these transformations, informs one that the vacuum expectation value of these fields must vanish.]

A symmetry is realized in a Nambu-Goldstone way if there exist some scalar
field (which may not necessarily be elementary) with non-zero vacuum expectation
value.  Imagine that this is so in Eq. (43).  Then using the definition of
the generators $G_i$ (Eq. (26)) one has
\begin{equation}
0 \not= -(g_i)_{\alpha\beta}\langle 0|\chi_\beta(x)|0\rangle =
\int d^3y\langle 0|J_i^o(y) \chi_\alpha(x) - \chi_\alpha(x)
J_i^o(y)|0\rangle~.
\end{equation}
This equation can be written in a more interesting way by inserting a complete
set of states $|n\rangle$ and making use of translational invariance on the
currents $J^o_i(y)$
\begin{equation}
J_i^o(y) = e^{-iPy}J_i^o(0)e^{iPy}~.
\end{equation}
Then the RHS of Eq. (46) reads
\begin{eqnarray}
{\rm RHS} &=& \sum_n \int d^3y\left\{\langle 0|e^{-iPy}J_i^o(0)e^{iPy}|n\rangle\langle n|\chi_\alpha(x)|0\rangle\right. 
\nonumber \\
& & \mbox{}-\left.\langle 0|\chi_\alpha(x)|n\rangle
\langle n|e^{-iPy}J_i^o(0) e^{iPy}|0\rangle\right\} \nonumber \\
&=&\sum_n \int d^3y e^{iP_ny}\langle 0|J_i^o(0)|n\rangle
\langle n|\chi_\alpha(x)|0\rangle \nonumber \\
& & \mbox{} -\sum_n \int d^3y e^{-iP_ny}
\langle 0|\chi_\alpha(x)|n\rangle\langle n|J_i^o(0)|0\rangle \nonumber \\
&=& \sum_n (2\pi)^3 \delta^3(\vec p_n)
\left\{e^{-iP_n^oy^o}\langle 0|J_i^o(0)|n\rangle
\langle n|\chi_\alpha(x)|0\rangle \right. \nonumber \\
& & \mbox{} \left.-e^{+iP_n^oy^o}\langle 0|\chi_\alpha(x)|n\rangle
\langle n|J_i^o(0)|0\rangle \right\}~.
\end{eqnarray}
By assumption this expression does {\bf not} vanish and, furthermore, since
the LHS is independent of $y^o$ it must also be independent of $y^o$.  Clearly
this can only happen if in the theory there exist some massless one-particle
states $|n\rangle$ and only these states contribute to the sum in (48).  
These zero mass states are the Goldstone bosons.

It is not difficult to convince oneself that for each generator $G_i$ that
does not annihilate the vacuum there is a corresponding Goldstone boson 
(after all the action of $G_i$ on the vacuum must give some state--and these
states are associated with the Goldstone bosons!).  Let us write the Goldstone boson states as
$|p;j\rangle$, where $p^2=0$.  Then it follows that the matrix element of the
currents associated with the broken generators between the vacuum and these
states are non-vanishing:
\begin{equation}
\langle0|J^\mu_i(0)|p;j\rangle = if_j\delta_{ij} p^\mu
\end{equation}
where $f_j$ are some non-vanishing constants, which are related to the vacuum
expectation values of the fields $\chi_\alpha$.  Indeed, remembering that for
a one-particle state
\begin{equation}
\sum_n \equiv \int \frac{d^3p_n}{(2\pi)^32p_n^o}
\end{equation}
it follows from Eqs. (46) and (48) that
\begin{equation}
i(g_i)_{\alpha\beta}\langle 0|\chi_\beta(0)|0\rangle = \lim_{p^{\mu}\to 0} 
\frac{1}{2}\left[f_i\langle p;i|\chi_\alpha(0)|0\rangle +
f^*_i\langle 0|\chi_\alpha(0)|p;i\rangle\right]~.
\end{equation}

Because the Nambu-Goldstone realization of a symmetry is so much less
familiar, it is instructive to illustrate it with a very simple example.  For
these purposes consider the following Lagrangian density describing the
interaction of a complex scalar field $\phi$ with itself
\begin{equation}
{\cal{L}} = -\partial_\mu\phi^{\dagger}\partial^\mu\phi - \lambda\left(\phi^{\dagger}\phi -
\frac{1}{2} f\right)^2~.
\end{equation}
Obviously this theory is invariant under a $U(1)$ transformation
(phase transformation)
\begin{eqnarray}
\phi(x) &\to& \phi^\prime(x) = e^{i\alpha}\phi(x) \nonumber\\
\phi^{\dagger}(x) &\to& \phi^{\prime \dagger}(x) = e^{-i\alpha}\phi^{\dagger}(x)~.
\end{eqnarray}
The conserved current associated with this symmetry is easily constructed
from our general formula (21)
\begin{equation}
J^\mu = \frac{\partial{\cal{L}}}{\partial\partial_\mu\phi}\frac{1}{i} (1)\phi
+ \frac{\partial{\cal{L}}}{\partial\partial_\mu\phi^{\dagger}}\frac{1}{i} (-1)
\phi^{\dagger} = i\left[(\partial^\mu\phi^{\dagger})\phi - (\partial^\mu\phi)\phi^
{\dagger}\right]~.
\end{equation}
The corresponding generator
\begin{equation}
G = \int d^3x J^o = i\int d^3x\left[(\partial^o\phi^{\dagger})\phi -
(\partial^o\phi)\phi^{\dagger}\right]
\end{equation}
obeys the commutation relations (15)
\begin{eqnarray}
[G,\phi(x)] &=& -\phi(x)  \\ \nonumber
[G,\phi^{\dagger}(x)] &=& +\phi^{\dagger}(x) ~.
\end{eqnarray}

In a classical sense, the second term in the Lagrangian correspond to a
potential for the fields $\phi,\phi^{\dagger}$:
\begin{equation}
V(\phi,\phi^{\dagger}) = \lambda\left(\phi^{\dagger}\phi - \frac{1}{2} f\right)^2~.
\end{equation}
Obviously, to guarantee the positivity of the theory, one needs that $\lambda > 0$.  However, the physics is very
different depending on the sign of $f$.  If $f<0$ the potential has a unique
minimum at $\phi = \phi^{\dagger}=0$ and the theory is realized in a Wigner-Weyl way,
leading to a degenerate multiplet of massive states.
If $f>0$, on the other hand, the potential has an infinity of minima
characterized by the condition $\phi^{\dagger}\phi = \frac{1}{2} f$.  The theory is
realized in a Nambu-Goldstone way and there is both a massless and a massive
state in the theory.

Quantum mechanically, if $f<0$, it is sensible to expand the potential about
$\phi = 0$, since this is the minimum of the potential.  One has
\begin{equation}
V = \lambda\left(\phi^{\dagger}\phi - \frac{1}{2} f\right)^2 = \frac{1}{4}
\lambda f^2 - \lambda f\phi^{\dagger}\phi + \lambda(\phi^{\dagger}\phi)^2~.
\end{equation}
The quadratic term $-\lambda f\phi^{\dagger}\phi$, since $f<0$, is a perfectly good
mass term for the fields $\phi$ and $\phi^{\dagger}$ and one identifies
\begin{equation}
m^2_\phi = m^2_{\phi^{\dagger}} = -\lambda f > 0~.
\end{equation}
In this case, one has a degenerate multiplet of two charge-conjugate particles interacting via
the $\lambda(\phi^{\dagger}\phi)^2$ term.

If $f>0$, on the other hand, an expansion about $\phi=0$ makes no sense as
the potential has a local maximum.  The only sensible point to expand the
potential is about its minimum value which occurs at $\phi_{\rm min} =
\sqrt{\frac{f}{2}} e^{i\theta}$.  In fact since $f>0$ there is no way
that the quadratic term in $\phi^{\dagger}\phi$ can represent a mass term.

Quantum mechanically the non-zero value of $\phi_{\rm min}$ implies that
$\phi$ has a non-vanishing vacuum expectation value
\begin{equation}
\langle \theta|\phi(x)|\theta\rangle = \sqrt{\frac{f}{2}} e^{i\theta}~.
\end{equation}
The phase $\theta$, characterizing the vacuum state $ |\theta\rangle$, is in fact irrelevant and can be rotated away.  It is a
reflection of the non-uniqueness of the vacuum state of the theory.  Since
under a $U(1)$ transformation
\begin{equation}
U^{-1}(\alpha) \phi)(x) U(\alpha) = e^{i\alpha}\phi(x)
\end{equation}
it is clear that the expectation of $\phi(x)$ between the states
$U(-\theta)|\theta\rangle$ is purely real
\begin{equation}
\langle \theta|U^{-1}(-\theta)\phi(x) U(-\theta)|\theta\rangle =
 e^{-i\theta} e^{i\theta}\sqrt{\frac{f}{2}}= \sqrt{\frac{f}{2}}~.
\end{equation}
Obviously $U(-\theta)|\theta\rangle\equiv |0\rangle$ is just as good a vacuum as $|\theta\rangle$.

Without loss of generality we can set $\theta=0$ and expand $\phi$ as
\begin{equation}
\phi = \sqrt{\frac{f}{2}} + \chi
\end{equation}
where the quantum field $\chi$, by assumption,  has  a vanishing vacuum expectation value.  The 
potential in terms of $\chi$ reads
\begin{eqnarray}
V &=& \lambda\left(\phi^{\dagger}\phi - \frac{f}{2}\right)^2 =
\lambda\left(\chi^{\dagger}\chi + \sqrt{\frac{f}{2}}(\chi + \chi^{\dagger})\right)^2\\ \nonumber
&=&\frac{\lambda f}{2}(\chi + \chi^{\dagger})^2 + \sqrt{2f}
\lambda(\chi + \chi^{\dagger})\chi^{\dagger}\chi + \lambda^2(\chi^{\dagger}\chi)^2~.
\end{eqnarray}
Obviously, it appears that a linear combination of $\chi$ and $\chi^{\dagger}$ has
a mass, while its orthogonal combination is massless.  Let us write
\begin{equation}
\chi_+ = \frac{1}{\sqrt{2}} (\chi + \chi^{\dagger})~; ~~~
\chi_- = \frac{i}{\sqrt{2}} (\chi^{\dagger}-\chi)~.
\end{equation}
Then
\begin{equation}
m_+^2 = 2\lambda f > 0~; ~~ m_-^2 = 0~.
\end{equation}
Even though the Langragian (52) is $U(1)$ symmetric, this symmetry is not
reflected in the spectrum, when the theory is realized in the Nambu-Goldstone
manner!

The above identification of $\chi$ as the Goldstone boson field also follows
directly from the commutators (56).  Since $f$ is real by assumption, one has
\begin{equation}
\chi_- = \frac{i}{\sqrt{2}} (\chi^{\dagger}-\chi) = \frac{i}{\sqrt{2}} (\phi^{\dagger}-\phi)
\end{equation}
and hence
\begin{equation}
[G,\chi_-] = \frac{i}{\sqrt{2}} (\phi^{\dagger} + \phi) =
i\left[\sqrt{f} + \chi_+\right]~.
\end{equation}
Whence, taking expectation values, one obtains
\begin{equation}
\langle 0|[G,\chi_-]|0\rangle = i\sqrt{f}~.
\end{equation}
This equation clearly singles out $\chi_-$ as the Goldstone boson field.

If $|p\rangle$ is the state corresponding to this  Goldstone boson  then, neglecting non-linearities, one
expects
\begin{equation}
\langle 0|\chi_-(0)|p\rangle = 1
\end{equation}
Eq. (69) then gives, in the same approximation,
\begin{equation}
\langle 0|J^\mu(0)|p\rangle = i\sqrt{f} p^\mu~.
\end{equation}
The decay constant $f_i$ of Eq. (49) here is just $\sqrt{f}$ and is related to the vacuum expectation value of $\phi$, as expected from Eq. (51).  There is an alternative way to accomplish this
identification by using directly the current $J^\mu$ and rewriting it in 
terms of the fields $\chi_+$ and $\chi_-$.  One has

\begin{eqnarray}
J^\mu &=& i[(\partial^\mu\phi^{\dagger})\phi - (\partial^\mu\phi)\phi^{\dagger}]  \\ \nonumber
&=& i\left[(\partial^\mu\chi^{\dagger})\left(\sqrt{\frac{f}{2}}+\chi\right) -
(\partial^\mu\chi)\left(\sqrt{\frac{f}{2}} + \chi^{\dagger}\right)\right] \\ \nonumber
&=&i\sqrt{f}\frac{1}{\sqrt{2}} \partial^\mu(\chi^{\dagger}-\chi) +
i[(\partial^\mu\chi^{\dagger})\chi - (\partial^\mu\chi)\chi^{\dagger}] \\ \nonumber
&=&\sqrt{f} \partial^\mu\chi_- + \mbox{non-linear terms}
\end{eqnarray}
which directly implies (71).

To summarize, there are two ways in which symmetries $([H,U]=0)$ can be 
realized in nature.  If the vacuum state is unique $(U|0\rangle = |0\rangle)$,
then we have a Wigner-Weyl realization with degenerate particle multiplets.
If, on the other hand, the vacuum state is not unique $(U|0\rangle \not= |0\rangle)$, then we have a Nambu-Goldstone realization with a number of massless
excitations, one for each of the generators of the group which does not annihilate the vacuum.  In this latter case one often refers to the phenomena
as spontaneous symmetry breaking because, although the symmetry exists, it is
not reflected in the spectrum of the states of the theory.

\section{Local Symmetries in Field Theory}

In all the preceding discussion I have talked implicitly only about
{\bf global} symmetry transformations.  That is the parameters $\delta a_i$
were assumed to be independent of space-time.  Clearly in this case fields at
different space-time points are transformed all in the same way.  One may well
ask what happens if the group parameters are space-time dependent.  In this
case the fields $\chi_\alpha(x)$ and $\chi_\alpha(x^\prime)$ would be rotated
in a different way by the group transformation.  Transformations where this
happens are called {\bf local} symmetries, to distinguish them from the
 global case when $\delta a_i$ is $x$-independent.

Under a {\bf local} transformation one has
\begin{equation}
\chi_\alpha(x) \stackrel{a(x)}{\rightarrow} \chi^\prime_\alpha(x) =
{\cal{R}}_{\alpha\beta}(a(x))\chi_\beta(x)~.
\end{equation}
 Because ${\cal{R}}$ is now space-time
dependent, even though the action
\begin{equation}
W = \int d^4x {\cal{L}}(\partial_\mu\chi_\alpha, \chi_\alpha)
\end{equation}
was invariant under {\bf global} $G$ transformations, this action will fail
to be invariant under {\bf local} $G$ transformations.  Because of the 
kinetic energy terms, which depends on $\partial_\mu\chi_\alpha$, there will
be pieces in $W$ which are no longer invariant.  Indeed, it is easy to
identify what destroys the possibility of local invariance of the action.
Consider the transformation of the derivative term $\partial_\mu\chi_\alpha$ under local transformations.
One has
\begin{eqnarray}
\partial_\mu\chi_\alpha(x) \stackrel{a(x)}{\rightarrow} \partial_\mu
\chi^\prime_\alpha(x) &=& \partial_\mu [{\cal{R}}_{\alpha\beta}(a(x))
\chi_\beta(x)] \nonumber \\
&=& {\cal{R}}_{\alpha\beta}(a(x))\partial_\mu\chi_\beta(x) +
\partial_\mu{\cal{R}}_{\alpha\beta}(a(x))\chi_\beta(x)~.
\end{eqnarray}
The presence of the second term above destroys the local invariance of the
action.  However, one can compensate for the appearance of this term by adding to the, globally invariant, Lagrangian
additional fields (gauge fields) which cancel this contribution.  It is clear that to
make a Lagrangian locally invariant necessarily involves the introduction of 
more degrees of freedom in the theory.

 Before
giving a general prescription of how to make a globally invariant Lagrangian
locally invariant, it is useful to illustrate this procedure with a simple example.  Consider a free Dirac field with Lagrangian density
\begin{equation}
{\cal{L}} = -\bar\psi(x)\left(\gamma^\mu \frac{1}{i}\partial_\mu + m\right)
\psi(x)~.
\end{equation}
Clearly ${\cal{L}}$ is invariant under the $U(1)$ transformation
\begin{eqnarray}
\psi(x) &\to& \psi^\prime(x) = e^{i\alpha}\psi(x) \nonumber \\
\bar\psi(x) &\to& \bar\psi^\prime(x) = e^{-i\alpha}\bar\psi(x)~,
\end{eqnarray}
which leads to the associated current:
\begin{equation}
J^\mu(x) = \frac{\partial{\cal{L}}}{\partial\partial_\mu\psi(x)}
\frac{1}{i} (1) \psi(x) = \bar\psi(x)\gamma^\mu\psi(x)~.
\end{equation}
It is clear, however, that if $\alpha = \alpha(x)$ the Lagrangian 
(76) ceases to be invariant, since
\begin{equation}
\partial_\mu\psi(x)\to \partial_\mu\psi^\prime(x) =
e^{i\alpha(x)}\partial_\mu\psi(x) + i(\partial_\mu\alpha(x))
\psi(x) e^{i\alpha(x)}~.
\end{equation}
Thus
\begin{eqnarray}
{\cal{L}}(x)\stackrel{a(x)}{\rightarrow}{\cal{L}}^\prime(x) &=&
{\cal{L}}(x) - (\partial_\mu\alpha(x))\bar\psi(x)\gamma^\mu\psi(x) \nonumber \\
&=& {\cal{L}}(x) - J^\mu(x)\partial_\mu\alpha(x)~.
\end{eqnarray}

One may get rid of the additional contribution in (80) by augmenting
the Lagrangian (76) by an additional term
\begin{equation}
{\cal{L}}_{\rm extra} = e A^\mu(x) J_\mu(x)
\end{equation}
involving a vector field $A^\mu(x)$, which under a local $U(1)$ transformation
translates by an amount $\partial_\mu\alpha(x)$:
\begin{equation}
A^\mu(x) \stackrel{a(x)}{\rightarrow} A^{\prime\mu}(x)= A^\mu(x)+ \frac{1}{e} (\partial^\mu\alpha(x))~.
\end{equation}
Of course, if this field $A^\mu(x)$ is to have a dynamical role, and one
wants to preserve the local invariance, the kinetic energy term for $A^\mu(x)$
should also be invariant under (82).  This is easily accomplished by
introducing the {\bf field strengths}:
\begin{equation}
F^{\mu\nu}(x) = \partial^\mu A^\nu(x) - \partial^\nu A^\mu(x)
\end{equation}
which are clearly invariant under (82).  Hence, the total Lagrangian
\begin{eqnarray}
{\cal{L}} &=& -\bar\psi(x)\left(\gamma^\mu\frac{1}{i}\partial_\mu + m\right)
\psi(x) + eA^\mu(x)\bar\psi(x)\gamma_\mu\psi(x) \nonumber \\
& & \mbox{} -\frac{1}{4} F^{\mu\nu}(x)F_{\mu\nu}(x)
\end{eqnarray}
involving the additional {\bf gauge field} $A^\mu$ is locally $U(1)$
invariant:
\begin{equation}
{\cal{L}}(x) \stackrel{a(x)}{\rightarrow} {\cal{L}}^\prime(x) =
{\cal{L}}(x)
\end{equation}
when
\begin{eqnarray}
\psi(x) &\stackrel{a(x)}{\rightarrow}& \psi^\prime(x) = e^{i\alpha(x)}\psi(x) \nonumber \\
A^\mu(x) &\stackrel{a(x)}{\rightarrow}& A^{\prime\mu}(x) = A^\mu(x) + \frac{1}{e}
\partial^\mu\alpha(x)~.
\end{eqnarray}

Note that to make the Lagrangian (76) locally $U(1)$ invariant it was
necessary to introduce an interaction term between the gauge fields $A^\mu$
and the globally conserved $U(1)$ current $J_\mu$.  There is a more geometrical
way to see how the interaction (81) is necessary to guarantee local
invariance.  As (79) demonstrates, the reason that the original Lagrangian
(76) is not locally invariant is because the derivative of the $\psi$ field
transforms {\bf inhomogeneously} under a local $U(1)$ rotation.  If one could
construct a modified derivative, $D_\mu\psi$, which under local transformations
transformed in the same way that $\partial_\mu\psi$ transforms under global
transformations, then the original Lagrangian could be trivially made locally
invariant by the replacement
\begin{equation}
{\cal{L}} (\partial_\mu\psi,\psi) \to {\cal{L}}(D_\mu\psi,\psi)~.
\end{equation}
Using Eq. (82), it is clear that for the case in question this modified
derivative--- a, so called, {\bf covariant derivative}--- is
\begin{equation}
D_\mu\psi = \partial_\mu\psi-ieA_\mu\psi~,
\end{equation}
since
\begin{eqnarray}
D_\mu\psi \stackrel{a(x)}{\rightarrow} D^\prime_\mu\psi^\prime &=& e^{i\alpha}\partial_\mu\psi
+ i(\partial_\mu\alpha)\psi e^{i\alpha}-ieA_\mu\psi e^{i\alpha} \nonumber \\
& & \mbox{} -i(\partial_\mu\alpha)\psi e^{i\alpha} \nonumber \\
&=& e^{i\alpha(x)}[\partial_\mu\psi-ie A_\mu\psi] = e^{i\alpha(x)}
D_\mu\psi~.
\end{eqnarray}
Obviously
\begin{equation}
{\cal{L}} = -\bar\psi(x)\left(\gamma^\mu\frac{1}{i} D_\mu + m\right)\psi(x)
-\frac{1}{4} F^{\mu\nu} F_{\mu\nu}
\end{equation}
is locally $U(1)$ invariant and coincides with the expression (84).

Viewed from this perspective, the demand of local invariance of a Lagrangian
is a marvelous prescription to fix the interactions of the globally invariant
fields with the gauge fields.  Furthermore, the gauge transformation (82)
does not allow the introduction of a mass term for the $A^\mu$ field, since
\begin{equation}
{\cal{L}}_{\rm mass} = -\frac{1}{2} m^2_A A^\mu(x) A_\mu(x)
\end{equation}
breaks the local $U(1)$ transformation.  So local invariance of a theory
severly restricts the dynamics.  In the example in question, it will be
recognized that the demand that a Dirac field be described by a Lagrangian
that is locally $U(1)$ invariant has produced the QED Lagrangian!  To
guarantee local $U(1)$ transformations it is necessary to introduce a massless
gauge field $A^\mu(x)$--- the photon field-- interacting with strength e--- the electric charge--- with the conserved current $J^{\mu}$.

The above simple example can be generalized to theories where the global
symmetry group is bigger than the $U(1)$ phase symmetry,~\cite{YM}
where the structure constants vanish (Abelian group).  For these purposes, consider again
a Lagrangian density ${\cal{L}}(\partial_\mu\chi_\alpha, \chi_\alpha)$
composed of fields which transform irreducibly under a non-Abelian group $G$
(a group where the structure constants $c_{ijk}\not= 0)$. Under {\bf global} $G$ transformations, one has
\begin{eqnarray}
\chi_\alpha(x) &\stackrel{a}{\rightarrow}& \chi^\prime_\alpha(x) =
{\cal{R}}_{\alpha\beta}(a)\chi_\beta(x)  \\ \nonumber
\partial_\mu\chi_\alpha(x) &\stackrel{a}{\rightarrow}& \partial_\mu
\chi^\prime_\alpha(x) = {\cal{R}}_{\alpha\beta}(a)
\partial_\mu\chi_\beta(x)~,
\end{eqnarray}
If this Lagrangian
density is invariant under these transformations then 
\begin{equation}
{\cal{L}}(\partial_\mu\chi_\alpha, \chi_\alpha) \stackrel{a}{\rightarrow}
{\cal{L}}^\prime(\partial_\mu\chi^\prime_\alpha, \chi^\prime_\alpha) =
{\cal{L}}(\partial_\mu\chi_\alpha, \chi_\alpha)~.
\end{equation}
Suppose one were able to introduce appropriate gauges fields to construct a {\bf covariant derivative}, $D_\mu\chi_\alpha(x)$,
which under {\bf local} $G$ transformations transformed as
$\partial_\mu\chi_\alpha(x)$ does under {\bf global} transformations. That is,
\begin{equation}
D_\mu\chi_\alpha(x) \stackrel{a(x)}{\rightarrow} D^\prime_\mu\chi^\prime_\alpha
(x) = {\cal{R}}_{\alpha\beta}(a(x))D_\mu\chi_\beta(x)~.
\end{equation}
Then, clearly, the Lagrangian ${\cal{L}}(D_\mu\chi_\alpha, \chi_\alpha)$ would
be {\bf locally} $G$ invariant
\begin{equation}
{\cal{L}}(D_\mu\chi_\alpha, \chi_\alpha)\stackrel{a(x)}{\rightarrow}
{\cal{L}}^\prime(D^\prime_\mu\chi^\prime_\alpha, \chi^\prime_\alpha) ={\cal{L}}(D_\mu\chi_\alpha, \chi_\alpha)~.
\end{equation}
For the theory to be physical, in addition, of course, one must also provide appropriate locally invariant field
strengths for the gauge fields entering in  the covariant derivatives
$D_\mu\chi_\alpha$.

By assumption, the  covariant derivatives required must transform under local
transformations as $\partial_\mu\chi_\alpha$  does in Eq. (92).  In analogy to what was done for the
simple $U(1)$ example, it is suggestive to introduce one gauge field
$A^\mu_i$ for each of the parameters $\delta a_i$ of the group $G$.  After
all, the gauge fields are supposed to compensate for the local variations of
the fields $\chi_\alpha$, and so there should be a gauge field for each of
the parameters $\delta a_i(x)$ of the Lie group G.  Taking the field
$\chi_\alpha(x)$ transformations under $G$ to be those of Eq. (15)
\begin{equation}
[G_i,\chi_\alpha(x)] = -(g_i)_{\alpha\beta}\chi_\beta
\end{equation}
the $U(1)$ example suggest writing for the covariant derivative $D_\mu\chi_\alpha$ the expression
\begin{equation}
D_\mu\chi_\alpha(x) = \left[\delta_{\alpha\beta}\partial_\mu - ig(g_i)_{\alpha\beta}A_{\mu i}(x)\right]\chi_\beta(x)~,
\end{equation}
where $g$ is some coupling constant.

For Eq. (92) to be satisfied for $D_\mu\chi_\alpha$, the gauge fields must respond appropriately
under local transformations.  To determine what this behavior should be, let
us compute $D^\prime_\mu\chi^\prime_\alpha$ and compare it to what we expect from (92).  One has
\begin{eqnarray}
D_\mu^\prime\chi^\prime_\alpha(x) &=& \partial_\mu\chi^\prime_\alpha(x) -
ig(g_i)_{\alpha\beta} A^\prime_{\mu i}(x)\chi^\prime_\beta(x) \nonumber \\
&=& \partial_\mu[{\cal{R}}_{\alpha\beta}(a(x))\chi_\beta(x)] - ig(g_i)_{\alpha\beta} A^\prime_{\mu i}(x){\cal{R}}_{\beta\gamma}(a(x))
\chi_\gamma(x) \nonumber \\
&=& {\cal{R}}_{\alpha\beta}(a(x))\partial_\mu\chi_\beta(x) +
\left(\partial_\mu{\cal{R}}_{\alpha\gamma}(a(x)\right)\chi_\gamma(x) \nonumber\\
& & \mbox{}-ig(g_i)_{\alpha\beta}A^\prime_{\mu i}(x)
{\cal{R}}_{\beta\gamma}(a(x))\chi_\gamma(x)~.
\end{eqnarray}
By definition we want
\begin{eqnarray}
D^\prime_\mu\chi^\prime_\alpha(x) &=& {\cal{R}}_{\alpha\beta}(a(x))
D_\mu\chi_\beta(x) \nonumber \\
&=&{\cal{R}}_{\alpha\beta}(a(x))\partial_\mu\chi_\beta - ig
{\cal{R}}_{\alpha\beta}(a(x))(g_i)_{\beta\gamma}
A_{\mu i}(x)\chi_\gamma(x)~.
\end{eqnarray}
It follows, therefore, that one must require that
\begin{equation}
-ig(g_i)_{\alpha\beta}A^\prime_{\mu i}(x){\cal{R}}_{\beta\gamma}(a(x)) +
\partial_\mu{\cal{R}}_{\alpha\gamma}(a(x)) = -ig{\cal{R}}_{\alpha\beta}(a(x))
(g_i)_{\beta\gamma}A_{\mu i}(x)~.
\end{equation}
Multiplying the above by ${\cal{R}}^{-1}$ finally gives the transformation required for the gauge field:
\begin{eqnarray}
(g_i)_{\alpha\beta}A^\prime_{\mu i}(x) &=& \frac{1}{ig}
[\partial_\mu{\cal{R}}_{\alpha\gamma}(a(x))]
[{\cal{R}}^{-1}(a(x))]_{\gamma\beta} \nonumber \\
& & \mbox{} + {\cal{R}}_{\alpha\gamma}(a(x))(g_i)_{\gamma\delta}
[{\cal{R}}^{-1}(a(x))]_{\delta\beta}A_{\mu i}(x)~.
\end{eqnarray}

It is easy to check that this formula agrees with Eq. (82) in the Abelian
$U(1)$ case when ${\cal{R}} = e^{i\alpha},~g_i=1$ and $g=e$.  In principle,
however, Eq. (101) has a very troublesome aspect, since it appears that the
transformation properties of the gauge fields $A^\prime_{\mu i}$ depend on
how the field $\chi_\alpha$ transforms under $G$.  If this were to be really the
case it would be disastrous, because to obtain a locally invariant theory one
would need to introduce a separate compensating gauge field for each matter
field in the theory.  Fortunately, although (101) as written appears to depend
on ${\cal{R}}$ explicitly, this dependence is in fact illusory.
{\bf The transformation properties of gauge fields depend only on the group
$G$ and not on how the matter fields transform.}

To prove this very important point, it is useful to consider Eq. (101) for
infinitesimal transformations, where
\begin{equation}
{\cal{R}}_{\alpha\beta}(\delta a(x)) = \delta_{\alpha\beta} + i~
\delta a_i(g_i)_{\alpha\beta}~.
\end{equation}
Using the above in (101), and employing an obvious matrix notation, one has
\begin{eqnarray}
g_kA^\prime_{\mu k}(x) &=& \frac{1}{ig}[\partial_\mu
(1+i\delta a_k(x)g_k)][1-i\delta a_i(x)g_i] \nonumber \\
& & \mbox{} +[1+i\delta a_j(x)g_j]g_i[1-i\delta a_k(x)g_k] 
A_{\mu i}(x) \nonumber \\
&\simeq& g_k A_{\mu k}(x) + i\delta a_j(x)[g_j,g_i]
A_{\mu i}(x) + \frac{1}{g}[\partial_\mu\delta a_k(x)] g_k~.
\end{eqnarray}
Using the commutation relations for the matrices $g_i$
\begin{equation}
[g_j,g_i] = ic_{jik}g_k = -ic_{ijk}g_k
\end{equation}
it is easy to see that the RHS of (103) is simply proportional to $g_k$
\begin{equation}
{\rm RHS} = g_k\left[A_{\mu k}(x) + c_{ijk}\delta a_j(x) A_{\mu i}(x) +
\frac{1}{g}[\partial_\mu\delta a_k(x)]\right]~.
\end{equation}
Thus, as anticipated, the transformation properties of the gauge fields are {\bf independent}
of the representation matrices $g_k$ associated with the fields $\chi_\alpha(x)$ and depend only on the structure constants of the group
$c_{ijk}$:
\begin{equation}
A^\prime_{\mu k}(x) = A_{\mu k}(x) + \delta a_j(x) c_{ijk}A_{\mu i}(x) +
\frac{1}{g}\partial_\mu(\delta a_k(x))~,
\end{equation}

For {\bf global transformations}, where the parameters $\delta a_k$ are
$x$-independent, the last term in (106) does not contribute and the
transformatiion of the gauge fields can be written in the standard form one
expects for a quantum field:
\begin{equation}
A^\prime_{\mu k}(x) =A_{\mu k}(x)+ i\delta a_j({\tilde g}_j)_{ki} A_{\mu i}(x)~. 
\end{equation}
Here the ``generator" matrices appropriate for the gauge fields, $\tilde g$,
are expressible in terms of the structure constants of the group
\begin{equation}
({\tilde g}_j)_{ki} = -ic_{ijk} = -ic_{jki}~.
\end{equation}
It is not hard to show (by using the Jacobi identity for $\tilde g_i,\tilde g_j,$ and $\tilde g_k$)
that the matrices ${\tilde g}$ in Eq. (108) indeed obey the group algebra of $G$
\begin{equation}
[{\tilde g}_i,{\tilde g}_j] = ic_{ijk}{\tilde g}_k~.
\end{equation}

The above discussion makes it clear that the gauge fields $A^\mu_i$ introduced in the covariant derivative (97) transform according to a special 
representation of the group $G$, the {\bf adjoint} representation.  If $G$
has $n$ parameters, then the matrices ${\tilde g}_i$ are $n\times n$
matrices, whose elements are related to the structure constants $c_{ijk}$.
Their transformation has no connection with how the matter fields $\chi_\alpha$
transform, but is intimately connected with the key parameters of $G$, its
structure constants.

Having made ${\cal{L}}$ locally invariant through the replacement of derivatives
by covariant derivatives, it remains to construct the field strengths for
the fields $A^\mu_i$, so as to be able to incorporate into the theory the
kinetic energy terms for the gauge fields.  It is easy to check that the naive
generalization of the Abelian example
\begin{equation}
{\tilde F}^{\mu\nu}_k = \partial^\mu A^\nu_k - \partial^\nu A^\mu_k
\end{equation}
will not work, since its transformation will still contain derivatives of the
parameters $\delta a_i$.  Indeed, using Eq. (106) one sees that
\begin{eqnarray}
{\tilde F}^{\prime\mu\nu}_k &=& \partial^\mu A^{\prime\nu}_k - \partial^\nu
A^{\prime\mu}_k = {\tilde F}_k^{\mu\nu} + \delta a_\ell c_{i\ell k}
{\tilde F}_i^{\mu\nu} \nonumber \\
& & \mbox{} + c_{ijk}\left[(\partial^\mu\delta a_j)A^\nu_i -
(\partial^\nu\delta a_j) A_i^\mu\right]~.
\end{eqnarray}
What one wants to do to obtain the correct field strengths is to augment (110) so as to eliminate altogether the
last term in (111).  Since this term contains both $\partial^\mu\delta a_j$
and $A^\nu_i$ in an antisymmetric fashion, one is led, after a bit
of reflection, to try the following ansatz for the non-Abelian field strengths:
\begin{equation}
F^{\mu\nu}_k(x) = \partial^\mu A^\nu_k(x)  -\partial^\nu A^\mu_k(x) + gc_{kij}
A_i^\mu(x) A_j^\nu(x)~.
\end{equation}

Let us check that Eq. (112) has the right properties. Using (106), the third term in (112) transforms as
\begin{eqnarray}
gc_{kij}A^\mu_i(x) A^\nu_j(x)& \to &gc_{kij}A^{\prime\mu}_i(x)
A^{\prime\nu}_j(x) \nonumber \\
&=& gc_{kij}\left[A^\mu_i(x) + \delta a_\ell(x) c_{m\ell i}A^\mu_m(x) +
\frac{1}{g}\partial^\mu\delta a_i(x)\right]\cdot \nonumber \\
& &\mbox{}\cdot\left[A^\nu_j + \delta a_\ell(x) c_{m\ell j}
A^\nu_m(x) + \frac{1}{g}\partial^\nu\delta a_j(x) \right] \nonumber \\
&\simeq& gc_{kij} A_i^\mu(x)A^\nu_j(x) + c_{kij}
\left[(\partial^\mu\delta a_i)A_j^\nu + 
(\partial^\nu\delta a_j) A_i^\mu\right] \nonumber \\
& &\mbox{} + \delta a_\ell(x)\left[gc_{kij}c_{m\ell i} A_m^\mu A^\nu_j +
gc_{kij}c_{m\ell j} A_i^\mu A^\nu_m\right].
\end{eqnarray}
However, making use of the antisymmetry of the structure constants, one has:
\begin{eqnarray}
c_{kij}(\partial^\mu\delta a_i)A_j^\nu &=& c_{kji}(\partial^\mu\delta a_j)
A_i^\nu = -c_{ijk}(\partial^\mu\delta a_j) A_i^\nu \nonumber \\
c_{kij}(\partial^\nu\delta a_j)A_i^\mu& = &c_{ijk}(\partial^\nu\delta a_j)
A_i^\mu~, 
\end{eqnarray}
and one sees that the last term in (111) precisely cancels the second term in
(113).

It is also not hard to check that the last term in (113) can be written in
a much more interesting form by making use of (109). Relabeling dummy indices and using (109) one obtains
\begin{eqnarray}
{\rm 3rd term} &=& g\delta a_\ell\left[c_{kij}c_{m\ell i}A_m^\mu A_j^\nu
+ c_{kmi}c_{j\ell i}A_m^\mu A^\nu_j\right]  \nonumber \\
&=& g\delta a_\ell\left[-c_{jki}c_{mi\ell} + c_{mki}c_{ji\ell}\right]
A_m^\mu A_j^\nu  \nonumber \\
&=& g\delta a_\ell[\tilde g_j,\tilde g_m]_{k\ell}
A_m^\mu A_j^\nu  \nonumber \\
&=& ig\delta a_\ell c_{jmp}[\tilde g_p]_{k\ell}A_m^\mu A_j^\nu \nonumber \\
&=& g\delta a_\ell c_{jmp}c_{pk\ell} A_m^\mu A_j^\nu \nonumber \\
&=& \delta a_\ell c_{i\ell k}\left[gc_{imj}A_m^\mu A_j^\nu\right]~.
\end{eqnarray}
Using the above one sees that what remains of (113)
transforms in precisely the same way as the second term of $\tilde F^{\mu\nu}_k$ [cf. Eq. (111)].  

Putting everything together, one sees that
under a local transformation the field strength $F_k^{\mu\nu}$ transforms
as
\begin{equation}
F_k^{\mu\nu}(x) \stackrel{\delta a(x)}{\rightarrow}
F_k^{\prime\mu\nu}(x) = F_k^{\mu\nu}(x) + \delta a_j(x) c_{ijk}
F_i^{\mu\nu}(x)~.
\end{equation}
The above is the desired result. Namely, that under local transformations the field strengths should transform as a quantum field  which belongs to the 
adjoint representation of the group.
In view of (116), it is easy to show that $F^{\mu\nu}_kF_{k\mu\nu}$ is
$G$-invariant.  One has
\begin{eqnarray}
F_k^{\mu\nu}F_{k\mu\nu} \to F^{\prime\mu\nu}_kF^\prime_{k\mu\nu} &=&
\left(F^{\mu\nu}_k + \delta a_j c_{ijk}F_i^{\mu\nu}\right)
\left(F_{k\mu\nu}+\delta a_jc_{ijk}F_{i\mu\nu}\right) \nonumber \\
&=& F_k^{\mu\nu}F_{k\mu\nu} + \delta a_j\left(c_{ijk}F_i^{\mu\nu}
F_{k\mu\nu} + c_{ijk}F_k^{\mu\nu}F_{i\mu\nu}\right) \nonumber \\
&=& F_k^{\mu\nu} F_{k\mu\nu}~,
\end{eqnarray}
since the 2nd term vanishes because of the antisymmetry of $c_{ijk}$:  
$c_{ijk} = -c_{jik}$. 

Let us recapitulate our results. The Lagrangian density ${\cal{L}}(
\partial_\mu\chi_\alpha, \chi_\alpha)$--- assumed to be invariant under {\bf global} $G$
transformations--- can be made locally invariant by introducing gauge
fields $A_i^\mu$, which enter in the covariant derivatives $D_\mu\chi_\alpha$ and the
field strengths $F_i^{\mu\nu}$.  The locally invariant Lagrangian density
is simply:
\begin{equation}
{\cal{L}}_{\rm local} = {\cal{L}}(D_\mu\chi_\alpha, \chi_\alpha) -
\frac{1}{4} F_i^{\mu\nu} F_{\mu\nu i}
\end{equation}
and is {\bf completely} determined from a knowledge of the global invariant
Lagrangian ${\cal{L}}$

Three remarks are in order:

\begin{description}
\item{i)} Again, as in the Abelian case, no mass term for the gauge fields
$A_i^\mu$ are allowed if one wants to preserve the local invariance
(106).
\item{ii)} The pure gauge Lagrangian
\begin{equation}
{\cal{L}} = -\frac{1}{4} F_i^{\mu\nu} F_{\mu\nu i}
\end{equation}
which contains the kinetic energy terms for the gauge fields $A_i^\mu$
is already a {\bf nonlinear} field theory, since $F_i^{\mu\nu}$ contains terms quadratic in the gauge fields $A_i^\mu$.  For the Abelian
case, where the structure constants vanish, these nonlinear terms are
absent.
\item{iii)} Because the gauge fields transform {\bf nontrivially} under
the group $G$, as far as global transformations go, the symmetry
currents of the full theory given by Eq. (118) now also get a contribution
from the gauge fields.  That is, one has
\begin{equation}
J_i^\mu = \frac{\partial {\cal{L}}}{\partial\partial_\mu\chi_\alpha}
\frac{1}{i} (g_i)_{\alpha\beta}\chi_\beta + \frac{\partial {\cal{L}}}
{\partial\partial_\mu A^\nu_j} \frac{1}{i} (\tilde g_i)_{jk} A_{\nu k}~.
\end{equation}
\end{description}

\section{The Higgs Mechanism}

We saw earlier that in the case of global symmetries, these symmetries could
be realized either in a Wigner-Weyl or Nambu-Goldstone way, depending on
whether the vacuum state was left, or not left, invariant by the group
transformations.  It is clearly of interest to know what happens in each of
these cases when the global symmetry is made local, via the introduction of
gauge fields.  For the Wigner-Weyl case, nothing very much happens.  Besides
the various degenerate multiplets of particles of the global symmetry there
is now also a degenerate zero mass multiplet of gauge field excitations.
In the Nambu-Goldstone case, however, some remarkable things happen.  When
the global symmetry is gauged, the Goldstone bosons associated with the
broken generators disappear and the corresponding gauge fields acquire a mass!
This is the celebrated Higgs mechanism. ~\cite{Higgs}

To understand this phenomena, it is useful to return to the simple $U(1)$
model discussed earlier and see what obtains when one tries to make the $U(1)$
global symmetry also a local symmetry of the Lagrangian.  Recall that the 
Lagrangian density of the model was
\begin{equation}
{\cal{L}} = -\partial_\mu\phi^\dagger\partial^\mu\phi - \lambda\left(\phi^\dagger\phi -
\frac{1}{2} f\right)^2~,
\end{equation}
and that the sign of $f$ determined whether one had a Wigner-Weyl realization
$(f<0)$ or a Nambu-Goldstone realization $(f>0)$.  To make the above
Lagrangian locally $U(1)$ invariant it suffices to replace $\partial_\mu\phi$
by a covariant derivative $D_\mu\phi$ involving a gauge field $A_\mu$, and
include in the theory a kinetic energy term for this gauge field.

If under local $U(1)$ transformations one assumes that
\begin{eqnarray}
\phi(x) &\to& \phi^\prime(x) = e^{i\alpha(x)}\phi(x) \\ \nonumber
A_\mu(x) &\to& A^\prime_\mu(x) = A_\mu(x) + \frac{1}{g}\partial_\mu\alpha(x)~,
\end{eqnarray}
then the covariant derivative
\begin{equation}
D_\mu\phi(x) = (\partial_\mu-igA_\mu)\phi
\end{equation}
clearly transforms just like $\phi$ does
\begin{equation}
D_\mu\phi(x) \to D^\prime_\mu\phi^\prime(x) = e^{i\alpha(x)}
(D_\mu\phi(x))~.
\end{equation}
Whence the augmented Lagrangian
\begin{equation}
{\cal{L}} = -(D_\mu\phi)^\dagger(D^\mu\phi) - \lambda\left(\phi^\dagger\phi -
\frac{1}{2} f\right)^2 - \frac{1}{4} F^{\mu\nu}F_{\mu\nu}
\end{equation}
with
\begin{equation}
F_{\mu\nu} = \partial_\mu A_\nu - \partial_\nu A_\mu
\end{equation}
is clearly locally $U(1)$ invariant.

If $f<0$, so that the global symmetry is Wigner-Weyl realized, the above
Lagrangian is suitable for computation as is.  It describes the interaction
of a degenerate multiplet of scalar fields ($\phi$ and $\phi^\dagger$) both with
themselves and with a massless gauge field $A_\mu$.  These latter interactions--since the $\phi$'s are scalar fields and hence have quadratic
kinetic energies--contain both a linear term in the gauge fields:
\begin{equation}
{\cal{L}}^{(1)}_{\rm int} = gA_\mu\left[i(\partial^\mu\phi^\dagger)\phi -
i\phi^\dagger\partial^\mu\phi\right] = gA_\mu J^\mu
\end{equation}
as well as a quadratic--so called ``sea-gull" term--contribution:
\begin{equation}
{\cal{L}}^{(1)}_{\rm sea~gull} = -g^2A^\mu A_\mu\phi^\dagger\phi~.
\end{equation}
These interactions follow directly from the gauge invariant replacement 
$\partial_\mu\phi \to D_\mu\phi = (\partial_\mu-igA_\mu)\phi$.

If $f>0$, on the other hand, so that the global $U(1)$ symmetry is realized
in a Nambu-Goldstone way, one must reparametrize the theory in terms of fields
with vanishing expectation value (c.f. Eq. (63)).  This reparametrization
is such that one is computing oscillations around the minimum of the
potential $V(\phi) $.
That is, one replaces
\begin{equation}
\phi^\dagger\phi = \frac{f}{2} + \mbox{quantum fields}~.
\end{equation}
This necessary shift implies that the seagull term of Eq. (128) gives rise to
a mass term for the $A_\mu$ field!
\begin{equation}
{\cal{L}}_{\rm mass} = -g^2 \frac{f}{2} A^\mu A_\mu \equiv -\frac{1}{2}
m_A^2 A^\mu A_\mu~.
\end{equation}
If the gauge field acquires mass, it follows that it cannot be purely
transverse (like the photon) but must also have a longitudinal polarization
component.  This extra degree of freedom must come from somewhere.  It is
not difficult to show that it arises from the {\bf dissapearance} of the
Nambu-Goldstone excitation, which would ordinarily arise from the spontaneous
breakdown of the global $U(1)$ symmetry.

To check this assertion, it is convenient to reparametrize the field
$\phi$, in the case $f>0$, in a somewhat different way than that chosen
before.  [The physics of the theory is, in fact, independent of the
parametrization one chooses, but certain parametrizations are more directly
physical.  Different choices for $\phi$ are akin to choosing different gauges
for $A_\mu$.]  Let us write $\phi$ in the following exponential parametrization:
\begin{equation}
\phi(x) = \frac{1}{\sqrt{2}}[\sqrt{f} + \rho(x)]\exp\left[i\frac{\xi(x)}
{\sqrt{f}}\right]~.
\end{equation}
Here $\rho(x)$ and $\xi(x)$ are real fields, with $\xi(x)$--the phase field--being connected to the Goldstone boson.  This last assertion is easy to understand
since $\xi(x)$ vanishes altogether from the potential $V$, and so obviously
cannot have any mass term.  One has simply
\begin{equation}
V = \lambda\left(\phi^\dagger\phi - \frac{f}{2}\right)^2 = \lambda
\left(\frac{\rho^2}{2} + \sqrt{f} \rho\right)^2~,
\end{equation}
so that the $\rho$ field has a mass
\begin{equation}
m^2_\rho = 2\lambda f
\end{equation}
in agreement with the value obtained earlier(cf. (66)).

It is easy to check that the phase field $\xi$ enters in the covariant
derivative in a trivial way, so that it can also be eliminated from the
kinetic energy term by an appropriate gauge choice.  Thus, as advertised,
the Nambu-Goldstone boson plays no role in the local theory.  It is
``eaten" to give mass to the gauge fields.  To prove this assertion, let us
consider $D_\mu\phi$ when $\phi$ is
 parametrized as in Eq. (131):
 \begin{eqnarray}
D_\mu\phi &=& (\partial_\mu-igA_\mu)\phi = (\partial_\mu-igA_\mu)
\frac{1}{\sqrt{2}} (\sqrt{f} + \rho) \exp\left[i\frac{\xi}{\sqrt{f}}\right] \\ \nonumber
&=&\frac{\exp\left[i\frac{\xi}{\sqrt{f}}\right]}{\sqrt{2}}\left[
\partial_\mu\rho -ig(\sqrt{f}+\rho)\left\{A_\mu - \frac{1}{g\sqrt{f}}
\partial_\mu\xi\right\}\right]~.
\end{eqnarray}
Obviously the factor in front of the $[~]$ bracket in (134) involving
$\exp\left[i\xi/\sqrt{f}\right]$ will not appear in the Lagrangian
(125), since the Lagrangian involves $(D_\mu\phi)^\dagger(D^\mu\phi)$.  Furthermore the $\xi$
dependence in the curly bracket is also spurious, since it can be eliminated
via a gauge transformation of the field $A^\mu$
\begin{equation}
A^\mu \to B^\mu = A^\mu - \frac{1}{g} \partial^\mu \frac{\xi}{\sqrt{f}}~.
\end{equation}

If the $U(1)$ global symmetry is spontaneously broken $(f>0)$ the Lagrangian
(125) can be rewritten entirely in terms of a {\bf massive} vector field
$B^\mu$ and a massive real scalar field $\rho$.  The resulting Lagrangian
\begin{eqnarray}
{\cal{L}} &=& -\frac{1}{2} \partial_\mu\rho\partial^\mu\rho - \frac{1}{2}
m^2_\rho \rho^2 - \frac{1}{4} F^{\mu\nu}F_{\mu\nu} - \frac{1}{2} m^2_A
B^\mu B_\mu \\ \nonumber
& & \mbox{} -g^2\left(\sqrt{f}\rho + \frac{1}{2} \rho^2\right)
B^\mu B_\mu - \lambda\left(\sqrt{f}\rho^3 + \frac{1}{4}\rho^4\right)
\end{eqnarray}
where
\begin{equation}
m^2_\rho = 2\lambda f~; ~~~ m_A^2 = g^2f
\end{equation}
shows no {\bf explicit} traces of the original $U(1)$ symmetry, except that
certain parameters in the interactions have particular interrelations. I remark that, although we demonstrated the absorption of the Goldstone boson to produce
a massive gauge field only in the Abelian case, this same phenomenon also
occurs in the non-Abelian case.

Let me close this section by discussing the two versions of the model
[Wigner-Weyl $f<0$; Nambu-Goldstone $f>0$] in terms of the degrees of freedom
present in the theory.  In the Wigner-Weyl case the theory has a complex
scalar field $\phi$ (2 degrees of freedom) plus a massless gauge field
$A^\mu$ (2 degrees of freedom, corresponding to the two transverse
polarizations).  In the Nambu-Goldstone case in the theory there is a real
scalar field $\rho$ (1 degree of freedom) plus a massive spin 1 field
$B_\mu$ (3 degrees of freedom).  Clearly both versions of the theory have the
same number of degrees of freedom.  However the spectrum of the excitations
is completely different!

\section{The Structure of Quantum Chromodynamics}

As a first illustration, I want to describe very briefly the structure of
Quantum Chromodynamics (QCD), the theory that describes the strong
interactions.~\cite{QCD}  As we shall see, although QCD is a local gauge theory
realized in a Wigner-Weyl way, it possesses also a set of approximate global
symmetries.  It turns out that some of these global symmetries are realized
in a Wigner-Weyl way, while some others are realized in a Nambu-Goldstone
manner.  Hence, QCD provides a nice practical example of the more formal
considerations we have discussed up to now.

We know in nature of the existence of six different types--{\bf flavors}--of
quarks: $u,d,s,c,b$, and $t$.  Each flavor of quark is actually a triplet of
fields, since the quarks transform irreducibly under the $SU(3)$ symmetry
group that characterizes QCD.  This $SU(3)$ symmetry is a local symmetry, so
besides quarks in QCD one must introduce the $3^2-1 = 8$ gauge fields which
are associated with the local $SU(3)$ symmetry.  These 8 gauge fields are known
as {\bf gluons}, since they help bind quarks into hadrons--- like protons and
$\pi$-mesons.

Let $q_\alpha^f(x)$ stand for a quark field, with the index $f$ denoting the
various flavors $f=\{u,d,s,c,b,t\}$ and $\alpha = \{1,2,3\}$ being an $SU(3)$
index.  Under local infinitesimal $SU(3)$ transformation then one has:
\begin{equation}
q_\alpha^f(x)\to q_\alpha^{\prime f}(x) =
\left[\delta_{\alpha\beta} + i\delta a_i(x)\left(\frac{\lambda_i}{2}\right)_{\alpha\beta}\right]
q_\beta^f(x)~.
\end{equation}
In the above, the $\lambda_i$ matrices $i=1,\ldots,8$ are the $3\times 3$
Gell-Mann matrices~\cite{GN} transforming as the {\bf 3} representation of $SU(3)$.
The $SU(3)$ structure constants--denoted here by $f_{ijk}$--are easily found by
using the explicit form of the $\lambda$-matrices given below:
\begin{equation}
\begin{array}{ccc}
\lambda_1 = \left[\begin{array}{ccc}
0 & 1 & 0 \\ 1 & 0 & 0 \\ 0 & 0 & 0 
\end{array}\right]~; &
\lambda_2 = \left[\begin{array}{ccc}
0 & -i & 0 \\ i & 0 & 0 \\ 0 & 0 & 0
\end{array} \right]~; &
\lambda_3 = \left[\begin{array}{ccc}
1 & 0 & 0 \\ 0 & -1 & 0 \\ 0 & 0 & 0
\end{array} \right]~;
\\
& & \\

\lambda_4 = \left[\begin{array}{ccc}
0 & 0 & 1 \\ 0 & 0 & 0 \\ 1 & 0 & 0
\end{array} \right]~; &

\lambda_5 = \left[\begin{array}{ccc}
0 & 0 & -i \\ 0 & 0 & 0 \\ i & 0 & 0
\end{array} \right]~; &

\lambda_6 = \left[\begin{array}{ccc}
0 & 0 & 0 \\ 0 & 0 & 1 \\ 0 & 1 & 0
\end{array} \right]~; \\
& & \\

\lambda_7 = \left[\begin{array}{ccc}
0 & 0 & 0 \\ 0 & 0 & -i \\ 0 & i & 0
\end{array} \right]~; &

\lambda_8 = \frac{1}{\sqrt{3}}\left[\begin{array}{ccc}
1 & 0 & 0 \\ 0 & 1 & 0 \\ 0 & 0 & -2
\end{array} \right]~. & \\
\end{array}
\end{equation}
One has
\begin{equation}
\left[\frac{\lambda_i}{2}~,~~\frac{\lambda_j}{2}\right] =
i~f_{ijk}\frac{\lambda_k}{2}~.
\end{equation}

The gauge fields $A^\mu_k(x)$ under a local infinitesimal $SU(3)$ transformation rotate into each other with coefficients proportional to the structure
functions $f_{ijk}$ and shift by the gradient of the
$SU(3)$ parameters $\delta a_k(x)$:
\begin{equation}
A_k^\mu(x)\to A^{\prime\mu}_k(x) =
A_k^\mu(x) + \delta a_jf_{ijk}A^\mu_i(x) +\frac{1}{g}
\partial^\mu\delta a_k(x)~.
\end{equation}
The field strengths
\begin{equation}
F_i^{\mu\nu} = \partial^\mu A_i^\nu - \partial^\nu A_i^\mu +
gf_{ijk}A_j^\mu A_k^\nu
\end{equation}
transform in the same way as the $A_k^\mu$ fields do but have {\bf no}
inhomogeneous contribution proportional to derivatives of $\delta a_k(x)$.  Finally,
the covariant derivatives of the quark fields
\begin{equation}
D^\mu_{\alpha\beta} q_\beta^f = \left[\partial^\mu\delta_{\alpha\beta} -
ig\left(\frac{\lambda_i}{2}\right)_{\alpha\beta} A_i^\mu\right]
q_\beta^f
\end{equation}
transform under local $SU(3)$ transformations precisely as the quark fields
themselves do.

Using the above equations, it is easy to see that the QCD Lagrangian
\begin{equation}
{\cal{L}}_{\rm QCD} = \sum_f - \bar q_\alpha^f
\left[\gamma^\mu\frac{1}{i}(D_\mu)_{\alpha\beta} + m_f\delta_{\alpha\beta}
\right]
q_\beta^f - \frac{1}{4} F_i^{\mu\nu}F_{i\mu\nu}
\end{equation}
is locally $SU(3)$ invariant.  In the above, the parameters $m_f$ are mass
terms for each flavor $f$ of quarks.  If these terms were absent, that is
if one could set $m_f\to 0$, it is clear that the QCD Lagrangian has a large
global symmetry in which quarks of one flavor are changed into quarks of
another flavor.  For six flavors of quarks, it is not difficult to show that,
in the limit $m_f\to 0$, the QCD Lagrangian is invariant under a $U(6)\times
U(6)$ group of global transformations.

Physically, it turns out that whether one can, or one cannot, approximately neglect the
quark mass terms $m_f$ depends on whether the mass $m_f$ is much smaller, or much greater,
than the dynamical scale, $\Lambda_{\rm QCD}$, associated with QCD. This latter scale is of order 300 MeV which is, in fact, much greater than
both the $u$- and $d$-quark masses. Although these, so called light quarks have masses much smaller than
$\Lambda_{\rm QCD}$:
\begin{equation}
m_{u,d} \ll \Lambda_{\rm QCD}~,
\end{equation}
it turns out that $m_s \sim \Lambda_{\rm QCD}$, while $\Lambda_{\rm QCD}$ is much smaller
than the masses of the $c$-, $b$- and $t$-quarks.  For this reason, in what follows, I will consider only the QCD piece of the Lagrangian involving
the $u$- and $d$-light quarks.  This, of course, is particularly interesting since these
quarks are the ones that make up ordinary hadrons, like the proton, neutron
and the pions.

The QCD Lagrangian for this 2-flavor case,  if we neglect for the moment
altogether $m_u$ and $m_d$, reads
\begin{equation}
\left.{\cal{L}}^{\rm 2-flavor}_{\rm QCD}\right|_{m_u=m_d=0} =
-\bar u_\alpha\gamma^\mu\frac{1}{i}[D_\mu]_{\alpha\beta}u_\beta -
\bar d_\alpha\gamma^\mu\frac{1}{i}[D_\mu]_{\alpha\beta}d_\beta -
\frac{1}{4} F_i^{\mu\nu}F_{i\mu\nu}~.
\end{equation}
Let us organize the $u$- and $d$-quarks into a doublet
\begin{equation}
Q_\alpha = \left(\begin{array}{c}
u \\ d 
\end{array} \right)_\alpha~,
\end{equation}
then the above Lagrangian can be written simply as
\begin{equation}
\left.{\cal{L}}^{\rm 2-flavor}_{\rm QCD}\right|_{m_u=m_d=0} =
-\bar Q_\alpha\gamma^\mu\frac{1}{i}[D_\mu]_{\alpha\beta} Q_\beta -
\frac{1}{4} F_i^{\mu\nu} F_{i\mu\nu}~.
\end{equation}
It is easy to check that this Lagrangian is invariant under a global
$U(2)_V\times U(2)_A$ symmetry in which
\begin{eqnarray}
Q \stackrel {V}{\rightarrow} Q^\prime = \exp\left[i~\alpha_i^VT_i\right] Q
&;& ~~\bar Q \stackrel {V}{\rightarrow} \bar Q^\prime = \bar Q
\exp\left[-i~\alpha_i^VT_i\right] \nonumber \\
Q \stackrel {A}{\rightarrow} Q^\prime = 
\exp\left[i~\alpha_i^AT_i\gamma_5\right] Q &;&
~~\bar Q^\prime \stackrel{A}{\rightarrow} \bar Q^\prime =
\bar Q\exp\left[i~\alpha_i^AT_i\gamma_5\right]~,
\end{eqnarray}
where the four $2\times 2$ matrices $T_i$ are just
\begin{equation}
T_i = \{\tau_i,1\}~,
\end{equation}
with $\tau_i$ the Pauli matrices.  The invariance under $U(2)_V$ is trivial to
see.  That under $U(2)_A$ follows once one realizes that  the Dirac matrix $\gamma_5$ anticommutes with all the $\gamma$-matrices: $\{\gamma_5,\gamma^\mu\} = 0$.

I should remark that even in the case when one restores the light quark masses in the Lagrangian, $m_u \not= 0,m_d\not= 0$, the QCD
Lagrangian is invariant under a common phase transformation of the $u$- and
$d$-quark fields.  This invariance just corresponds to the phase invariance
associated with the overall quark number, or baryon number, with $U(1)_V \subset
U(2)_V$.  If $m_u=m_d$, then it is easy to show that also the remaining
$SU(2)$ subgroup in $U(2)_V$,  $SU(2)_V \subset
U(2)_V$, is conserved.  This subgroup is just the usual
isospin, well known from nuclear physics.~\cite{Heisenberg}  Notice, however, that isospin is an
approximate symmetry of QCD even if $m_u \not= m_d$, provided that the
absolute value of these masses is much less than $\Lambda_{\rm QCD}$.  In
this case, to a good approximation one can neglect both $m_u$ and $m_d$
(even if they are not equal!) and the strong interactions are then invariant
under the $SU(2)$ isospin group.

The approximate $U(2)_V\times U(2)_A$ invariance of the strong interactions
was discovered in the 1960's even before QCD was put forth as the theory of the
strong interactions.~\cite{AD}  It was realized then, however, that while the
$U(2)_V$ global symmetry appeared to be realized in nature as a Wigner-Weyl
symmetry, the $U(2)_A$ symmetry was realized in a Nambu-Goldstone way. 
Indeed, if $U(2)_A$ were an approximate Wigner-Weyl symmetry, not only would one
expect a degenerate neutron-proton doublet but also one should have another
doublet of states, of opposite parity,  approximately degenerate with the neutron-proton
doublet.  Because this additional degenerate doublet was not seen in the
spectrum of baryons, this approximate $U(2)_A$ symmetry
must be spontaneously broken.  In this case, one would expect some (nearly) massless
Nambu-Goldstone states to appear in the theory.  The triplet of pions
$(\pi^+,\pi^-,\pi^o)$, which are much lighter than any other meson states,
were suggested as the likely candidate for these approximately Nambu-Goldstone states. Indeed, one can show that, dynamically, these states really behave as approximate Nambu-Goldstone states should. For instance, at low energy their couplings vanish linearly with energy.

Matters were clarified further with the advent of QCD, since one was able to understand
better  both the origin of the approximate symmetry and the mechanism which causes the breakdown of $U(2)_V\times U(2)_A$. Let me briefly comment on this last point. In 
QCD, because of the same strong forces that confine quarks into hadrons,
condensates of $u$- and $d$-quarks can form.  These condensates are
nothing but non-zero expectation values of quark bilinears in the QCD vacuum.
Clearly if
\begin{equation}
\langle 0|\bar u(0) u(0)|0\rangle =
\langle 0|\bar d(0) d(0)|0\rangle \not= 0~,
\end{equation}
although $U(2)_V\times U(2)_A$ is an (approximate) symmetry of the QCD
Lagrangian, only $U(2)_V$ remains as a true symmetry of the spectrum.  That is,
the above condensates breaks
\begin{equation}
U(2)_V\times U(2)_A \to U(2)_V
\end{equation}

Naively one would expect as a result of the above spontaneous breakdown that
four Nambu-Goldstone bosons should appear in the theory.  In fact, the
$U(1)_A$ subgroup of the $U(2)_A$ group, although it is a symmetry at the Lagrangian level, can be shown not to be a real
{\bf quantum} symmetry of QCD.~\cite{chiral} Radiative effects cause the divergence of the $U(1)_A$ current not to vanish. Unfortunately, the argument why the $U(1)_A$ symmetry acquires an anomalous divergence--- a, so called, {\bf chiral anomaly}~\cite{ABJ}--- 
is too complex to enter upon here.  Nevertheless, taking this result
at face value, one expects that the formation of the condensates above should  produce
3 Nambu-Goldstone bosons, associated with the breakdown of the $SU(2)_A$
symmetry.  These states are the pions.  Indeed, one can show that the pion
mass attains a finite value once one turns on the $u$- and $d$-quark masses,
but vanishes in the limit as $m_u$, $m_d\to 0$.~\cite{DH}  I will not pursue this
point further here, but note only how simply one can understand the
approximate symmetry properties of the strong interactions, deduced in the 1960s  after much hard work,~\cite{GOR} directly from the QCD Lagrangian and a few dynamical assumptions, Eqs. (145) and (151).

\section{The Structure of the $SU(2)\times U(1)$ Theory}

The ideas we have just discussed of a spontaneously broken gauge theory have
found a spectacular application in the $SU(2)\times U(1)$ model of the
electroweak interactions of Glashow, Salam and Weinberg.~\cite{GSW}  At first sight, it
appears that weak and electromagnetic interactions have little in common,
so that a combined gauge model of these forces does not appear very natural.
However, there were at least two phenomenological similarities which hinted at
a common link, and which helped in the formulation of the $SU(2)\times U(1)$
model.

The first of these similarities is that in both weak and electromagnetic interactions {\bf currents} are involved.
In the electromagnetic case the interaction Lagrangian
\begin{equation}
{\cal{L}}_{\rm em} = eA^\mu J_\mu^{\rm em}
\end{equation}
gives rise to long-range forces between charged particles due to the exchange of
a massless photon field.  The $1/r$ potential between charged particles 
follows from the $1/q^2$ propagator for the photon field.  The effective
action among charged particles due to (153) is simply
\begin{eqnarray}
W_{\rm em}^{\rm eff} &=& \frac{i}{2} \int d^4x~ eJ^\mu_{\rm em}(x)
\langle T(A_\mu(x)A_\nu(y))\rangle d^4y~ eJ_{\rm em}^\nu(y) \\ \nonumber
&=& \frac{1}{2}\int d^4x~ eJ^\mu_{\rm em}(x)
D_{\mu\nu}(x-y) d^4y ~eJ^\nu_{\rm em}~,
\end{eqnarray}
where $D_{\mu\nu}$ is the photon propagator.  Since the currents
$J_\mu^{\rm em}$ are conserved, one can take effectively
\begin{equation}
D_{\mu\nu}(x-y) = \eta_{\mu\nu} \int \frac{d^4q}{(2\pi)^4} e^{iq(x-y)}
\frac{1}{q^2-i\epsilon}~,
\end{equation}
where
\begin{equation}
\eta_{\mu\nu} = \left(
\begin{array}{cccc}
-1 &   &   &   \\
   & 1 &   &   \\
   &   & 1 &   \\
   &   &   & 1
\end{array} \right) \nonumber
\end{equation}
is the metric tensor.  Hence
\begin{eqnarray}
W_{\rm eff}^{\rm em} &=& \frac{1}{2} \int d^4x d^4y \frac{d^4q}{(2\pi)^4}
eJ^\mu_{\rm em}(x) e^{iq(x-y)}\frac{1}{q^2-i\epsilon} e J_\mu^{\rm em}(y) \\ \nonumber
&=&\frac{1}{2}\int \frac{d^4q}{(2\pi)^4}
\left[eJ^\mu_{\rm em}(q) \frac{1}{q^2-i\epsilon}
eJ_\mu^{\rm em}(-q)\right]~.
\end{eqnarray}
Thus, in momentum space, one has simply
\begin{equation}
{\cal{L}}_{\rm eff}^{\rm em}(q) 
= \frac{1}{2}\left[eJ^\mu_{\rm em}(q)\frac{1}{q^2} eJ_\mu^{\rm em}(-q)\right]~.
\end{equation}
\
For the charged current weak interactions, which are responsible for the
rather long lived nuclear disintegrations, like neutron $\beta$ decay, one
has known for a long time that they could be described by an effective
current-current theory, the Fermi theory:~\cite{Fermi}
\begin{equation}
{\cal{L}}_{\rm Fermi} = \frac{G_F}{\sqrt{2}} J_+^\mu(x) J_{-\mu}(x)~.
\end{equation}
Here $G_F$--the Fermi constant--has dimensions of ${\rm (mass)}^{-2}$
and $G_F \sim 10^{-5}$ (GeV)$^{-2}$.  In momentum space (159) looks like
the e.m. case, except that the photon propagator $1/q^2$ is replaced
by the constant $G_F/\sqrt{2}$. In momentum space, one has
\begin{equation}
{\cal{L}}^{\rm cc}_{\rm eff}(q) =
\frac{G_F}{\sqrt{2}} \left[J_+^\mu(q) J_{-\mu}(q)\right]~.
\end{equation}

This phenomenological resemblance can be sharpened by imagining that the
contact nature of the charged current weak interactions is due to the
exchange of a very heavy ``weak boson".  For low momentum transfer processes,
the propagator of the weak boson would be effectively constant
\begin{equation}
\frac{1}{q^2+M^2_W} \stackrel{q^2\ll M^2_W}{\simeq} \frac{1}{M^2_W}~.
\end{equation}
So Eq. (159) could arise from an interaction Lagrangian very similar to that
of electromagnetism:~\cite{Schwinger}
\begin{equation}
{\cal{L}}_{\rm weak} = \tilde g[J_+^\mu(x) W_{-\mu}(x) +
J_-^\mu(x) W_{+\mu}(x)]
\end{equation}
involving some spin one bosons $W_\pm^\mu$.  Then one could obtain, for
$q^2\ll M^2_W$, ${\cal{L}}^{\rm cc}_{\rm eff}$ from the exchange of these
massive fields.
\begin{equation}
{\cal{L}}^{\rm cc}_{\rm eff}(q) 
\stackrel{q^2\ll M^2_W}{\simeq} \frac{\tilde g^2}{M^2_W}\left[ J_+^\mu(q)
J_{-\mu}(-q)\right]~,
\end{equation}
which identifies the Fermi constant as
\begin{equation}
\frac{G_F}{\sqrt{2}} = \frac{\tilde g^2}{M^2_W}~.
\end{equation}
Note that if $\tilde g^2\sim e^2$ then from the value of $G_F$ one infers
that the masses of the weak bosons are really heavy: $M_W \sim 100$ GeV!

The second similarity between weak and electromagnetic processes is that the charged currents that enter in weak decays appear to be related
to the electromagnetic current--at least as far as the strongly interacting
particles go.  This interrelation was discussed long ago by Feynman and
Gell-Mann, and by Marshak and Sudarshan.~\cite{FGMS}  The vector piece of the $J_\pm^\mu$ 
currents are identical to the $1\mp i2$ components of the strong isospin
current.  In turn the isovector piece of the electromagnetic current is the
3rd component of this same strong isospin current.

Although the above two points hint at a possible common origin of weak and
electromagnetic interactions, they are not {\it per se} compelling.  The
dominant reason for attempting to treat both interactions on the same footing
is theoretical.  The Fermi theory (159) is actually a very sick theory as it
stands, since in higher order in perturbation theory one encounters 
divergences which one cannot eliminate from the theory.  These divergences occur because of the very singular nature of
the contact interaction (159) which, in contrast
to what happens in QED, is not being damped at all for large $q^2$.  

It turns out that matters are not ameliorated even
if the Fermi theory is replaced by an interaction like (162), involving
mediating heavy vector bosons $W_\pm^\mu$.  This is because the propagator for
such a massive boson contains in the numerator a propagator factor, 
characteristic of a spin one object, which is badly behaved at large $q^2$:
\begin{eqnarray}
\Delta_{\mu\nu}(q) &=& \frac{1}{q^2+M^2_W}
\left(\eta_{\mu\nu} + \frac{q_\mu q_\nu}{M^2_W} \right)\\ \nonumber
&\stackrel{q^\mu~{\rm large}}{\rightarrow}& O(1)~.
\end{eqnarray}
Thus it is not possible to add ``by hand" an interaction like (162) and hope 
to obtain a sensible weak interaction theory.  If, however, the interaction
(162) resulted from making a global symmetry local--- so that the $W_\pm^\mu$ are
gauge fields which are massive because of the Higgs mechanism --- then the
situation is vastly improved.  It turns out that the gauge invariance of the
theory allows one to calculate higher order corrections with propagators for
the $W$-fields which have only the $\eta_{\mu\nu}$ term.  These theories, as
first shown by 't Hooft,~\cite{tH} have the same {\bf good asymptotic behavior} as QED.
They are renormalizable.

The above argues for a theory of the weak interactions based on some symmetry
group $G$ which spontaneously breaks down.  Two of the currents associated
with $G$ must include $J_+^\mu$ and $J_-^\mu$.  However, the generator algebra
must close and so one expects naturally also some neutral current.  This
current, in general, will be related to the electromagnetic current.  Thus we
see that renormalizability has lead us directly to contemplate models in
which at the Lagrangian level, weak and electromagnetic currents enter on
the same footing!

The simplest unified model of the electroweak interactions, which contains
$J_+^\mu,~J_-^\mu$ and $J_{\rm em}^\mu$ is based on the group $O(3)$.~\cite{GG}  
However, the discovery of weak neutral current processes experimentally argued for at least a
4-parameter group.  The suggestion of Glashow, Salam and Weinberg, made well
before the discovery of these neutral currents  processes, was that the electroweak 
interactions are based on an $SU(2)\times U(1)$ gauge theory, which suffers
spontaneous breakdown to $U(1)_{\rm em}$.  This theory has three
massive gauge bosons, associated with the broken generators, and a massless
gauge field, associated with the photon.  The model was built to reproduce
the known structure of the charged current weak interactions.  It then predicted
particular neutral current interactions, whose experimental 
verification provided a direct test of the model.  Furthermore,
the model also predicts the masses of the gauge fields associated with the
spontaneous breakdown.  The observation at CERN of the $W^\pm$ and $Z^o$ bosons, ~\cite{WZ} with the masses predicted by the model, provided the final experimental confirmation of
the validity of the $SU(2)\times U(1)$ theory.

To detail the structure of the GSW model, one has to specify how the
matter degrees of freedom transform under the $SU(2)\times U(1)$ group.  This
could be deduced from the form of the charged currents $J_\pm^\mu$, which
a long series of experiments in the 1950's and 1960's showed to have a
(V-A) form.~\cite{LY}  That is, only the left-handed projection of the fermionic fields
appear to participate in these interactions.  For instance, from a study of $\beta$-decay for the
muon one established that the current $J_+^\mu$ had both $\mu-\nu_\mu$ and
$e-\nu_e$ terms, in which only the left-handed neutrino fields entered:
\begin{equation}
J_+^\mu = \bar e\gamma^\mu(1-\gamma_5)\nu_e + \bar\mu\gamma^\mu(1-\gamma_5)\nu_\mu
+ \ldots~.
\end{equation}
Writing the projections
\begin{equation}
\psi = \frac{1}{2}(1-\gamma_5)\psi + \frac{1}{2}(1+\gamma_5)\psi =
\psi_{\rm L} + \psi_{\rm R}
\end{equation}
and using the properties
\begin{equation}
\{\gamma_5,\gamma^\mu\} = 0~;~~\gamma_5^2 = 1~;~~\bar\psi = \psi^\dagger\gamma^o~;
~~\gamma_5^\dagger = \gamma_5~,
\end{equation}
one sees that
\begin{equation}
J_
+^\mu = 2\bar e_{\rm L}\gamma^\mu\nu_{e{\rm L}} + 2\bar\mu_{\rm L}
\gamma^\mu\nu_{\mu{\rm L}} + \ldots~.
\end{equation}
That is, the charged currents only contain left-handed fields.

The structure of $J_+^\mu$, and its complex conjugate $J_-^\mu$, suggests
that under $SU(2)$ the $\nu_{e{\rm L}}$ and the $e_{\rm L}$ fields
(and the $\nu_{\mu{\rm L}}$ and $\mu_{\rm L}$ fields) transform as a doublet.
The appropriate generator matrix for an $SU(2)$ doublet is $\frac{\tau_i}{2}$,
where $\tau_i$ are the Pauli matrices.  Indeed these matrices obey the
$SU(2)$ Lie algebra
\begin{equation}
\left[\frac{\tau_i}{2}~,~\frac{\tau_j}{2}\right] = i~\epsilon_{ijk}
\frac{\tau_k}{2}~.
\end{equation}
Hence, if $\left(\begin{array}{c}
\nu_e \\ e \end{array}
\right)_{\rm L}$ transforms as a doublet, the relevant piece of the $SU(2)$
current involving these fields is
\begin{equation}
J_i^\mu = (\bar\nu_e~\bar e)_{\rm L} \gamma^\mu
\frac{\tau_i}{2}
\left(\begin{array}{c}
\nu_e \\ e
\end{array} \right)_{\rm L}
\end{equation}
and one sees that indeed
\begin{eqnarray}
2(J_1^\mu - iJ_2^\mu) &=& 2(\bar\nu_e~\bar{ e})_{\rm L}
\gamma^\mu\left(
\begin{array}{cc}
0 & 0 \\ 1 & 0
\end{array} \right)
\left(\begin{array}{c}
\nu_e \\ e
\end{array} \right)_{\rm L} \\ \nonumber
&=& 2\bar e_{\rm L}\gamma^\mu\nu_{e{\rm L}} =
(J_+^\mu)_{\nu_e-e}
\end{eqnarray}

The fundamental matter entities presently known are quarks and leptons,
which appear in a repetitive pattern as far as the $SU(2)\times U(1)$
interactions are concerned [cf. the $\nu_e-e$ and $\nu_\mu-\mu$ terms in
$J_+^\mu$ of Eq. (166)].  To date we know of the existence of three
generations of quarks and leptons: the electron family: $(\nu_e,e;u,d)$; the muon family
$(\nu_\mu,\mu;c,s)$ and the $\tau$-lepton family $(\nu_\tau,\tau;t,b)$, where
to each lepton doublet there are associated a pair of quarks.  The quarks
in the pair actually are comprised each of three states, since each
quark carries a color index $\alpha=1,2,3$.  As we just discussed, these color degrees of freedom are associated with the strong interactions of quarks, which are
based on an $SU(3)$ gauge theory realized in a Wigner-Weyl way--- QCD.

Because all the three families transform in the same way under $SU(2)\times
U(1)$, I will only describe the $SU(2)\times U(1)$ properties of the electron
family.  In view of the preceding discussion, it is clear that
$\left(\begin{array}{c} 
\nu_e \\ e 
\end{array} \right)_{\rm L}$ transforms as an $SU(2)$ doublet.  So does the
quark pair $\left( \begin{array}{c}
u \\ d 
\end{array} \right)_{\rm L}$, as an analysis of beta decay of nuclei indicates.
Furthermore, since only left-handed fields enter in the weak charged currents,
it must be that the right-handed components of the electron family are
$SU(2)$ singlets.  Since the $SU(2)\times U(1)$ group must eventually break
down to $U(1)_{\rm em}$, it follows that the electromagnetic charge must be a
linear combination of the $U(1)$ generator and  of the neutral $T_3$ generator of $SU(2)$, which is diagonal. Thus one can write
\begin{equation}
Q = T_3 + Y~,
\end{equation}
with $Y$ being the $U(1)$ generator.  Hence the $U(1)$ quantum numbers of the
fields in the electron family follow from their known charges.  These 
considerations allow us to build the following table for the transformation
properties of $\nu_e,e,u$ and $d$ under $SU(2)\times U(1)$. The right-handed neutrino field $\nu_{\rm R}$ in Table 1 is usually not included as a
real excitation, since it is a {\bf total} $SU(2)\times U(1)$ singlet and so
does not participate in these interactions.

\begin{table}
\caption{Transformation properties of quarks and leptons}
\begin{center}
\begin{tabular}{c|c|c|c}
States & SU(2) & U(1) & Q \\ \hline
 & & & \\ 
 {$\left(\begin{array}{c}
\nu_e \\ e
\end{array} \right)_{\rm L}$} & 2 & -1/2 & {$\left(\begin{array}{c}
0 \\ -1 
\end{array} \right)$} \\  \hline
 & & & \\
{$\left(\begin{array}{c}
u \\ d
\end{array} \right)_{\rm L}$} & 2 & 1/6 &
{$\left(\begin{array}{c}
2/3 \\ -1/3
\end{array} \right)$} \\ \hline
 & & & \\
{$(\nu_e)_{\rm R}$} & 1 & 0 & 0 \\ \hline
 & & & \\ 
{$(e)_{\rm R}$} & 1 & -1 & -1 \\  \hline
 & & & \\
{$(u)_{\rm R}$} & 1 & 2/3 & 2/3 \\ \hline
 & & & \\
{$(d)_{\rm R}$} & 1 & -1/3 & -1/3 \\ \hline
 & & &
\end{tabular}
\end{center}
\end{table}

Given the transformation properties of the quarks and leptons under
$SU(2)\times U(1)$, we may now immediately write down the {\bf locally}
$SU(2)\times U(1)$ invariant Lagrangian which describes their interactions.
For that purpose we need only to replace in the {\bf free} Dirac Lagrangian
for the fermion fields the ordinary derivatives $\partial_\mu\psi$ by the
appropriate $SU(2)\times U(1)$ covariant derivatives $D_\mu\psi$ and add
the gauge field interactions.  Using Table 1, it is trivial to write down
these covariant derivatives.  One has
\begin{eqnarray}
D_\mu\left(\begin{array}{c}
\nu_e \\ e
\end{array} \right)_{\rm L} &=& 
\left(\partial_\mu - ig\frac{\tau_i}{2} W_{\mu i} + ig^\prime
\frac{1}{2} Y_\mu\right) \left(\begin{array}{c}
\nu_e \\ e
\end{array} \right)_{\rm L} \\
D_\mu\left(\begin{array}{c}
u \\ d
\end{array} \right)_{\rm L} &=&
\left(\partial_\mu - ig\frac{\tau_i}{2} W_{\mu i} - ig^\prime\frac{1}{6}
Y_\mu \right) \left( \begin{array}{c}
u \\ d
\end{array} \right)_{\rm L}  \\
D_\mu ~ \nu_{\rm R} &=& (\partial_\mu ) \nu_{\rm R} \\
D_\mu ~ e_{\rm R} &=& (\partial_\mu + ig^\prime Y_\mu) e_{\rm R} \\
D_\mu ~ u_{\rm R} &=& (\partial_\mu - ig^\prime \frac{2}{3}
Y_\mu) u_{\rm R} \\
D_\mu ~ d_{\rm R} &=& (\partial_\mu + ig^\prime \frac{1}{3}
Y_\mu) d_{\rm R}~.
\end{eqnarray}
Here $g,g^\prime$ are the $SU(2)$ and $U(1)$ coupling constants, respectively,
 while $W_{\mu i}$ and $Y_\mu$  are the $SU(2)$ and $U(1)$ gauge fields, respectively.

The Lagrangian for the $SU(2)\times U(1)$ model of Glashow, Salam and
Weinberg--- as far as the interactions among the fermions of the electron
family and the gauge fields go--- is then simply
\begin{eqnarray}
{\cal{L}}_{\rm FG} = &-& (\bar\nu_e~\bar e)_{\rm L} \gamma^\mu\frac{1}{i}
D_\mu \left(\begin{array}{c}
\nu_e \\ e
\end{array} \right)_{\rm L} - (\bar u~\bar d)_{\rm L}
\gamma^\mu \frac{1}{i} D_\mu\left(\begin{array}{c}
u \\ d
\end{array} \right)_{\rm L} \\ \nonumber
&-& \bar e_{\rm R} \gamma^\mu \frac{1}{i} D_\mu e_{\rm R} -
\bar u_{\rm R}\gamma^\mu\frac{1}{i} D_\mu e_{\rm R} -
\bar d_{\rm R} \gamma^\mu\frac{1}{i} D_\mu d_{\rm R} \\ \nonumber
&-& \frac{1}{4} W_i^{\mu\nu} W_{\mu\nu i} - \frac{1}{4}
Y^{\mu\nu} Y_{\mu\nu}~,
\end{eqnarray}
where the field strengths $W_i^{\mu\nu}$ and $Y^{\mu\nu}$ are given by
\begin{eqnarray}
W_i^{\mu\nu} &=& \partial^\mu W_i^\nu - \partial^\nu W_i^\mu +
g\epsilon_{ijk} W_j^\mu W_k^\nu \\
Y^{\mu\nu} &=& \partial^\mu Y^\nu - \partial^\nu Y^\mu~.
\end{eqnarray}
Note that the Lagrangians (180) contains {\bf no} mass terms for the
fermion fields.  Mass terms involve a left-right transition
\begin{equation}
{\cal{L}}_{\rm mass} = -m\bar\psi\psi = -m(\bar\psi_{\rm L}\psi_{\rm R} +
\bar\psi_{\rm R}\psi_{\rm L})~.
\end{equation}
Since under $SU(2)$ $\psi_{\rm L} \sim 2$ and $\psi_{\rm R} \sim 1$, clearly
the $SU(2)\times U(1)$ symmetry permits no fermion mass terms.  As I will
show later, however, masses can be generated when $SU(2)\times U(1)$ is
spontaneously broken down.

Before we discuss the breakdown of $SU(2)\times U(1)$ it is useful to organize
a bit the interaction terms which emerge from the Lagrangian (180).  These
take the simple form
\begin{equation}
{\cal{L}}_{\rm int} = gW_i^\mu J_{\mu i} + g^\prime Y^\mu J_{\mu Y}
\end{equation}
where the $SU(2)$ and $U(1)$ currents, $J_i^\mu$ and $J_Y^\mu$ are readily
seen to be
\begin{eqnarray}
J_i^\mu &=& (\bar\nu_e~\bar e)_{\rm L} \gamma^\mu\frac{\tau_i}{2}
\left(\begin{array}{c}
\nu_e \\ e
\end{array} \right)_{\rm L} +
(\bar u~\bar d)_{\rm L} \gamma^\mu\frac{\tau_i}{2}
\left(\begin{array}{c}
u \\ d
\end{array} \right)_{\rm L} \\
J_Y^\mu &=& -\frac{1}{2}(\bar\nu_e~\bar e)_{\rm L}\gamma^\mu
\left(\begin{array}{c}
\nu_e \\ e 
\end{array} \right)_{\rm L} +
\frac{1}{6} (\bar u~\bar d)_{\rm L} \gamma^\mu
\left(\begin{array}{c}
u \\ d
\end{array} \right)_{\rm L} \\ \nonumber
& & \mbox{} -\bar e_{\rm R}\gamma^\mu e_{\rm R} + \frac{2}{3}
\bar u_{\rm R}\gamma^\mu u_{\rm R} - \frac{1}{3}
\bar d_{\rm R}\gamma^\mu d_{\rm R}~.
\end{eqnarray}
I note that since in the model the electromagnetic current is given by
[cf. Eq. (173)]
\begin{equation}
J^\mu_{\rm em} = J^\mu_3 + J^\mu_Y
\end{equation}
the phenomenological observation mentioned earlier, that the vector piece of
the weak charged currents and the isovector piece of $J^\mu_{\rm em}$ are
related, is built in already in (187).

It is convenient to rewrite (184) in terms of physical fields.  If the
model is to reproduce the weak interactions, the $SU(2)\times U(1)$ symmetry
must suffer a spontaneous breakdown to $U(1)_{\rm em}$.  This means that of
the four gauge fields $W_i^\mu,~Y^\mu$, three must acquire a mass and one
will remain massless.  Now, in general, $U(1)_{\rm em}$  is a
linear combination of an $U(1) \subset SU(2)$ and $U(1)_Y$, so that
one expects the photon fields to be a linear combination of $W_3^\mu$ and
$Y^\mu$.  The orthogonal combination then corresponds to a massive neutral
field--- the $Z^o$ boson.  It has become conventional to parametrize these
linear combinations in terms of an angle $\theta_W$--- the Weinberg angle.
\begin{eqnarray}
W_3^\mu &=& \cos\theta_W Z^\mu + \sin\theta_W A^\mu \nonumber \\
Y^\mu &=& -\sin\theta_W Z^\mu + \cos\theta_W A^\mu
\end{eqnarray}
It proves useful also to rewrite $W_1^\mu$ and $W_2^\mu$ in terms of fields of
definite charge
\begin{equation}
W_\pm^\mu = \frac{1}{\sqrt{2}} (W_1^\mu \mp i~W_2^\mu)
\end{equation}
and use the charged currents $J_\pm^\mu$, which enter in the Fermi theory
[cf. Eq. (159)]
\begin{equation}
J_\pm^\mu = 2(J_1^\mu \mp i~J_2^\mu)~.
\end{equation}
With all these definitions the interaction Lagrangian of Eq. (184) becomes
\begin{eqnarray}
{\cal{L}}_{\rm int} &=& \frac{g}{2\sqrt{2}}
[W_+^\mu J_{-\mu} + W_-^\mu J_{+\mu}] \nonumber \\
& & \mbox{} \left\{(g\cos\theta_W + g^\prime\sin\theta_W)
J_3^\mu - g^\prime\sin\theta_W J^\mu_{\rm em}\right\} Z_\mu \nonumber \\
& & \mbox{} +\left\{g^\prime\cos\theta_W J^\mu_{\rm em} +
(g^\prime\cos\theta_W-g\sin\theta_W) J_3^\mu \right\} A_\mu~.
\end{eqnarray}
In the above, I have made use of (187) to eliminate altogether $J_Y^\mu$ in
favor of $J^\mu_{\rm em}$.

The above interaction is supposed to reproduce both the electromagnetic
interaction (153) and the charged current weak interaction (162). It predicts as well
a new neutral current weak  interaction involving the $Z^o$ boson.  Since the
photon field is supposed to only interact with $J^\mu_{\rm em}$ with strength $e$, one sees that one must
require the Weinberg angle to obey the {\bf unification} condition
\begin{equation}
g^\prime\cos\theta_W = g\sin\theta_W = e~.
\end{equation}
Using this information to eliminate $g$ and $g^\prime$ in terms of
$\theta_W$ and $e$  allows one to write for the interaction Lagrangian the expression:
\begin{eqnarray}
{\cal{L}}_{\rm int} &=& \frac{e}{2\sqrt{2}\sin\theta_W}
(W_+^\mu J_{\mu -} + W_-^\mu J_{\mu +}) + eJ^\mu_{\rm em} A_\mu \nonumber \\
& & \mbox{} + \frac{e}{2\cos\theta_W\sin\theta_W} J^\mu_{\rm NC} Z_\mu~.
\end{eqnarray}
Here the neutral current $J^\mu_{\rm NC}$ which interacts with the $Z_\mu$
field is
\begin{equation}
J^\mu_{\rm NC} = 2[J_3^\mu - \sin^2\theta_W J^\mu_{\rm em}]~.
\end{equation}
Comparing this result with our earlier discussion, the coupling $\tilde g$ of Eq. (162)  is seen to be
\begin{equation}
\tilde g = \frac{e}{2\sqrt{2}\sin\theta_W}~.
\end{equation}
Hence, the comparison with the Fermi theory [cf. Eq. (164)] gives
for the Fermi constant the expression
\begin{equation}
\frac{G_F}{\sqrt{2}} = \frac{\tilde g^2}{M^2_W} =\frac{e^2}{8\sin^2\theta_W M_W^2}~.
\end{equation}
One sees that a knowledge of the Weinberg angle--which enters in the neutral
current--gives direct information on the mass of the heavy weak boson which mediates the charged current weak interactions.  One finds experimentally that
$\sin^2\theta_W \simeq 1/4$,~\cite{Hollik} which predicts for $M_W$
a value of around 80 GeV.  This prediction has been spectacularly confirmed
by the discovery at the CERN Collider of a particle of this mass with all the
characteristic of the $W$ boson.

Just as charged current interactions, for processes where the momentum
transfer $q^2 \ll M^2_W$, can be described by the Fermi theory, one can arrive at a similar structure for neutral
current interactions.  In the same approximation,  $q^2 \ll M^2_Z$, one has
\begin{equation}
{\cal{L}}^{\rm NC}_{\rm Fermi} 
\simeq \frac{1}{2}\left[\frac{e}{2\sin\theta_W\cos\theta_W}\right]^2
\frac{1}{M^2_Z} J^\mu_{\rm NC} J_{\mu{\rm NC}}~.
\end{equation}
Using the identification (195) of the Fermi constant, one has
\begin{eqnarray}
{\cal{L}}_{\rm Fermi}^{\rm NC} &=& \frac{G_F}{\sqrt{2}}
\left[\frac{M^2_W}{M^2_Z\cos^2\theta_W}\right] J^\mu_{\rm NC} J_{\mu {\rm NC}}
\nonumber \\
&=& \frac{G_F}{\sqrt{2}} \rho J^\mu_{\rm NC} J_{\mu{\rm NC}}~,
\end{eqnarray}
where the ratio
\begin{equation}
\rho = \frac{M^2_W}{M^2_Z\cos^2\theta_W}
\end{equation}
gives the relative strength of neutral to charged current weak processes.

To summarize, the weak interactions in the Glashow Salam Weinberg model, in the
limit in which $q^2 \ll M^2_W,M^2_Z$ can be written in a current-current form
\begin{equation}
{\cal{L}}^{\rm eff}_{\rm Weak} = \frac{G_F}{\sqrt{2}}
[J_+^\mu J_{-\mu} + \rho J^\mu_{\rm NC} J_{\mu{\rm NC}}]~.
\end{equation}
The charged current weak interactions by construction agree with experiment.
Neutral current weak interactions test the model, since {\bf all} experiments
must be describable by the only two free parameters $\rho$ and $\sin^2\theta_W$, which enters in the definition of $J^\mu_{\rm NC}$, present in (194).  All neutral current experiments, indeed, can be fitted with a common value
of $\sin^2\theta_W \simeq 1/4$ and of $\rho \simeq 1$,~\cite{Hollik} thereby providing strong
confirmation of the validity of the GSW model.  Furthermore, given $\rho$ and
$\sin^2\theta_W$, one can determine the mass of the $Z^o$ and $W^\pm$ bosons from Eqs.
(196) and (199).  The discovery at the CERN collider of the $W^\pm$ bosons  and, soon thereafter, of a neutral heavy particle of mass around 90 GeV, in agreement with
the value predicted by the GSW model, provided a splendid confirmation of the
model.~\cite{WZ}

To complete the GSW model, it is
necessary to describe briefly the mechanism by which the $W^\pm$ and $Z^o$
bosons get mass.  The idea here is very much like that described in the last
section, when I discussed the Higgs mechanism for the Abelian $U(1)$ model.
Namely, one introduces some scalar field whose self interactions cause the
$SU(2)\times U(1)$ symmetry to break down.  Since we want $SU(2)$ to break
down, the scalar field introduced into the theory must carry $SU(2)$ quantum
numbers.  The simplest possibility is afforded by an $SU(2)$ doublet.  
Furthermore, since we want also to break the $U(1)$ symmetry, this doublet
must be complex.  Thus the simplest agent to carry through the desired
breakdown is the complex doublet.
\begin{equation}
\Phi = \left(\begin{array}{c}
\phi^o \\ \phi^-
\end{array} \right)~,
\end{equation}
where $\phi^o$ and $\phi^-$ are complex fields, and the charge assignments
identify $Y_\phi = -1/2$.

To accomplish the breakdown we consider a potential analogous to that in
(58).  In addition we must introduce an appropriately $SU(2)\times U(1)$
covariant kinetic energy term for the field $\Phi$, using the covariant
derivative
\begin{equation}
D_\mu \Phi = \left(\partial_\mu - ig\frac{\tau_i}{2} W_{\mu i} +
i\frac{g^\prime}{2} Y_\mu\right) \Phi.
\end{equation}
The interaction Lagrangian involving the scalar field $\Phi$--the Higgs
field--is just then:
\begin{equation}
{\cal{L}}_{\rm HG} = -(D_\mu\Phi)^\dagger (D^\mu\Phi) -
\lambda \left(\Phi^\dagger\Phi - \frac{v^2}{2}\right)^2~.
\end{equation}
It is clear that the potential term in (203) will cause $SU(2)\times U(1)$
to break down.  The choice of vacuum expectation value
\begin{equation}
\langle\Phi\rangle = \frac{v}{\sqrt{2}}
\left(\begin{array}{c}
1 \\ 0
\end{array} \right)
\end{equation}
guarantees that $SU(2)\times U(1)\to U(1)_{\rm em}$.  [Actually, with just one
doublet $\Phi$ one can always define $U(1)_{\rm em}$ as the $U(1)$ left
unbroken in $V$.  The choice (204) is dictated by our definition of charge.
Any other choice would do, but it would change what we called $Q$.]

Given the vacuum expectation value (204), the mass terms for the gauge
fields are read off immediately from the seagull terms:
\begin{equation}
{\cal{L}}_{\rm seagull} = -\left[\left(g\frac{\tau_i}{2} W_{\mu i}-g^\prime
\frac{1}{2} Y_\mu\right)\Phi\right]^\dagger
\left[\left(g\frac{\tau_i}{2} W_i^\mu - g^\prime\frac{1}{2} Y^\mu\right)\Phi\right]~.
\end{equation}
Using Eq. (188), the gauge field matrix in (205) is easily seen to be
\begin{eqnarray}
g\frac{\tau_i}{2} W_i^\mu - g^\prime\frac{1}{2} Y^\mu &=&
\left[\begin{array}{cc}
\frac{g}{2} W_3^\mu - \frac{g^\prime}{2} Y^\mu & \frac{g}{\sqrt{2}} W_+^\mu \\
\frac{g}{\sqrt{2}} W_-^\mu & -\frac{g}{2} W_3^\mu - \frac{g^\prime}{2}
Y^\mu 
\end{array} \right] \\ 
&=& \left[\begin{array}{cc}
\frac{g}{2\cos\theta_W} Z^\mu & \frac{g}{\sqrt{2}} W_+^\mu \\
\frac{g}{\sqrt{2}} W_-^\mu & \frac{g}{2\cos\theta_W}
[\sin^2\theta_W-\cos^2\theta_W]Z^\mu-A^\mu
\end{array} \right]~. \nonumber
\end{eqnarray}
Since the vacuum expectation value (204) only has an upper component,
one sees that {\bf only} $Z^\mu$ and not $A^\mu$ acquires a mass, confirming
our previous identification of this latter field as the photon field.
Replacing in (205) $\Phi\to\langle\Phi\rangle$ gives the following mass
terms for the gauge fields:
\begin{equation}
{\cal{L}}_{\rm mass} = -\left(\frac{gv}{2}\right)^2 W_+^\mu W_{-\mu} -
\frac{1}{2}\left(\frac{gv}{2\cos\theta_W}\right)^2 Z^\mu Z_\mu~.
\end{equation}
Hence
\begin{equation}
M^2_W = \frac{1}{4} (gv)^2~; ~~ M^2_Z = \frac{1}{4\cos^2\theta_W}
(gv)^2~.
\end{equation}
We see that the simplest choice of Higgs field to give the $SU(2)\times U(1)
\to U(1)_{\rm em}$ breaking predicts that the parameter $\rho$ in the neutral
current interactions is unity!
\begin{equation}
\rho = \frac{M^2_W}{M^2_Z\cos^2\theta_W} = 1~.
\end{equation}
The experimental indications that $\rho\simeq 1$ suggest therefore that nature
has chosen (again) the simplest course.  Using Eq. (208) and the relation
of $M_W$ to the Fermi constant identifies the scale parameter $v$ in the Higgs
potential as
\begin{equation}
v = (\sqrt{2}~{\rm G_F})^{-1/2} \simeq 250~{\rm GeV}~.
\end{equation}

The introduction of a doublet Higgs field $\Phi$ into the theory has another
salutary effect--it allows for the possibility of generating masses for the
quarks and leptons!  I will illustrate the idea with the up
 quark.  Since
$\Phi$ carries hypercharge $Y_\Phi = -1/2$ and is an $SU(2)$ doublet, the
interaction (Yukawa interaction) of $\Phi$ with $u_{\rm R}$ and the
$(\bar u~\bar d)_{\rm L}$ doublet is allowed by the $SU(2)\times U(1)$ symmetry:
\begin{equation}
{\cal{L}}_{\rm Yukawa} = -h(\bar u ~\bar d)_{\rm L}\Phi u_{\rm R}-h^*
\bar u_{\rm R}\Phi^\dagger \left(\begin{array}{c}u \\ d \end{array} \right)_{\rm L}~.
\end{equation}

Obviously, when $\Phi$ has a vacuum expectation value this interaction will
generate a mass term for the $u$ quark.  Taking $h$ real, one has
\begin{equation}
{\cal{L}}_{\rm mass} = -\frac{hv}{\sqrt{2}}\bar uu = - m_u\bar uu~,
\end{equation}
so that $m_u$ is also related to the breakdown parameter $v$.  Unfortunately
since $h$ is not known, no predictions follow.  This same mass generation
procedure holds for all quarks and leptons.

\section*{Acknowledgements}

I am extremely grateful to J. Tran Than Van for having invited me to lecture in the fifth Vietnam School of Physics. The extremely friendly atmosphere of the school and of all the participants made my stay in Hanoi a real pleasure. This 
work was supported in part by the department of energy under contract No. DE-FG03-91ER40662, Task C.


\section*{References}

\begin{thebibliography}{99}

\bibitem{Ramond} See, for example, P. Ramond, {\bf Field Theory: A Modern Primer}, (Addison Wesley, Redwood City, California, 1989).

\bibitem{Group} For a useful discussion of group theory in general, and Lie groups in particular, see, for example, W. K. Tung {\bf Group Theory in Physics}, (World Scientific, Singapore, 1985).

\bibitem{WW} These symmetries are appropriately named after Eugene Wigner and Hermann Weyl, who were the pioneers in the applicaton of symmetry groups in physics. Two illustrative papers by them are: E. P. Wigner, Proc. Nat. Acad. Sci. U. S. {\bf 38} (1952) 449; H. Weyl, Z. Phys. {\bf 56} (1929) 330.

\bibitem{NG} The idea of having a spontaneously broken symmetry originated in connection with the BCS theory of superconductivity. For early work on this see, for example, P. W. Anderson, Phys. Rev. {\bf 110} (1958) 827. The consequences of spontaneous breakdown of a symmetry are first explored in the seminal papers of Nambu and Goldstone: Y. Nambu, Phys. Rev. Lett. {\bf 4} (1960) 380; J. Goldstone, Nuovo Cimento {\bf 19} (1961) 154. These ideas are further elucidated in Y. Nambu and G. Jona-Lasinio, Phys. Rev. {\bf 122} (1961) 345; {\bf 124} (1961) 246; and in J. Goldstone, A. Salam and S. Weinberg, Phys. Rev. {\bf 127} (1962) 965.

\bibitem{LSZ} H. Lehmann, K. Symanzik and W. Zimmermann, Nuovo Cimento {\bf 1} (1955) 1425.

\bibitem{YM} C. N. Yang and R. L. Mills, Phys. Rev. {\bf 96} (1954) 191.

\bibitem{Higgs} P. W. Higgs, Phys. Lett. {\bf 12} (1964) 132; Phys. Rev. Lett. {\bf 13} (1964) 508; F. Englert and R. Brout, Phys. Rev. Lett. {\bf 13} (1964) 321: see also, G. S. Guralnik, C. Hagen and T. W. B. Kibble, Phys. Rev. Lett. {\bf 13} (1964) 585.

\bibitem{QCD} Many people contributed to the development of Quantum Chromodynamics (QCD). Some of the key references include:  H. Fritzsch, M. Gell-Mann and H. Leutwyler, Phys. Lett. {\bf 47B} (1973) 365; H. D. Politzer, Phys. Rev. Lett. {\bf 30} (1973) 1346; D. Gross and F. Wilczek, Phys. Rev. Lett. {\bf 30} (1973) 1343; Phys. Rev. {\bf D8} (1973) 3633; S. Weinberg, Phys. Rev. Lett. {\bf 31} (1973) 494.


\bibitem{GN} See, for example, M. Gell-Mann and Y. Neeman, {\bf The Eightfold Way} (W. A. Benjamin, Inc. New York, N. Y., 1964). 
 
\bibitem{Heisenberg} W. Heisenberg, Z. Phys. {\bf 77} (1932) 1.

\bibitem{AD} See, for example, S. L. Adler and R. F. Dashen, {\bf Current Algebras} ( W. A. Benjamin, Inc. New York, N. Y., 1968). 

\bibitem{chiral} G. 't Hooft, Phys. Rev. Lett. {\bf 37} (1976) 8.

\bibitem{ABJ} S. L. Adler, Phys. Rev. {\bf 177} (1969) 47, J.S. Bell and R. Jackiw, Nuovo Cimento {\bf 60} (1969) 47.

\bibitem{DH} For a full discussion see, for example, J. F. Donoghue, E. Golowich and B. R. Holstein, {\bf Dynamics of the Standard Model} (Cambridge University Press, Cambridge, U. K., 1994).

\bibitem{GOR} Two representative papers are: M. Gell-Mann, R. J. Oakes and B. Renner, Phys. Rev. {\bf 175} (1968) 2195; and R.F. Dashen, Phys. Rev. {\bf 183} (1969) 1245.

\bibitem{GSW} S. L. Glashow, Nucl. Phys. {\bf B22} (1961) 579; S. Weinberg, Phys. Rev. Lett. {\bf 19} (1967) 1264; A. Salam, in {\bf Elementary Particle Theory}, ed. N. Svartholm (Amquist and Wiksels, Stockholm, 1969).

\bibitem{Fermi} E. Fermi, Z. Phys. {\bf 88} (1934) 361.

\bibitem{Schwinger} J. Schwinger, Ann. Phys. {\bf 2} (1957) 407.

\bibitem{FGMS} R. P. Feynman and M. Gell-Mann, Phys. Rev. {\bf 109} (1958) 193; R. E. Marshak and E. C. G. Sudarshan, Phys. Rev. {\bf 109} (1958) 1860.

\bibitem{tH} G. 't Hooft, Nucl. Phys. {\bf B35} (1971) 167.

\bibitem{GG} H. Georgi and S. L. Glashow, Phys. Rev. Lett. {\bf 28} (1972) 1494.

\bibitem{WZ} UA1 Collaboration, G. Arnison {\it et. al}, Phys. Lett. {\bf 122B} (1983) 103; {\bf 126B} (1983) 398; UA2 Collaboration, M. Banner {\it et. al}, Phys. Lett. {\bf 122B} (1983) 476; P. Bagnaia {\it et. al}, Phys. Lett. {\bf 129B} (1983) 130. 

\bibitem{LY} The presence of an axial component in the weak charged current, leading to parity violation in weak decays, was suggested originally by T. D. Lee and C. N. Yang, Phys. Rev. {\bf 104} (1956) 254.

\bibitem{Hollik} For a discussion of the present status of tests of the GSW theory see, W. Hollik and G. Duckeck, {\bf Electroweak Precision Tests at LEP} (Springer Verlag, Berlin, 2000).


\end{thebibliography}
\end{document}